\documentclass[aps,pre,preprint,onecolumn,showpacs,superscriptaddress]{revtex4-1}
\usepackage{graphicx}
\usepackage{amsmath}
\usepackage{amssymb}
\usepackage{psfrag}
\usepackage{natbib}
\usepackage{color}

\newcommand{\be}{\begin{equation}}
\newcommand{\ee}{\end{equation}}

\newcommand{\bea}{\begin{eqnarray}}
\newcommand{\eea}{\end{eqnarray}}

\begin{document}
	\title{Dynamics of oscillators globally coupled via two mean fields}
\author{Xiyun Zhang}
\affiliation{Department of Physics, East China Normal University,
Shanghai, 200062, P. R. China}	
\affiliation{Institute for Physics and Astronomy,
University of Potsdam, Karl-Liebknecht-Str. 24/25, 14476 Potsdam-Golm, Germany}
\author{Arkady Pikovsky}
\affiliation{Institute for Physics and Astronomy,
University of Potsdam, Karl-Liebknecht-Str. 24/25, 14476 Potsdam-Golm, Germany}
\affiliation{Department of Control Theory, Nizhny Novgorod State University,
Gagarin Av. 23, 606950, Nizhny Novgorod, Russia}
\author{Zonghua Liu}
\affiliation{Department of Physics, East China Normal University,
Shanghai, 200062, P. R. China}
\begin{abstract}
Many studies of synchronization properties of coupled oscillators, based on the classical
Kuramoto approach, focus on ensembles coupled via a mean field.
Here we introduce a setup of Kuramoto-type phase oscillators coupled via two mean fields.
We derive stability properties of the incoherent state and find traveling wave solutions with different
locking patterns; stability properties of these waves are found numerically.
Mostly nontrivial states appear when the two fields compete, i.e. one tends to synchronize oscillators
while the other one desynchronizes them. Here we identify normal branches which bifurcate from the incoherent
state in a usual way, and anomalous branches, appearance of which cannot be described as a bifurcation. Furthermore,
hybrid branches combining properties of both are described. In the situations where no stable traveling wave exists,
modulated quasiperiodic in time dynamics is observed. Our results indicate that a competition between two coupling channels
can lead to a complex system behavior,  providing a potential generalized framework for understanding of complex phenomena in
natural oscillatory systems.
\end{abstract}

	\date{\today}

	\maketitle

Dynamics of globally coupled oscillators attracted large attention recently. It is relevant for many physical systems, like
Josephson junctions, lasers, arrays of spin-torque  and electronic oscillators~\cite{Cawthorne_etal-99,Nixon_etal-13,Temirbayev_etal-12,Grollier-Cros-Fert-06},
but also for many life and social
systems~\cite{Richard-Bakker-Teusink-Van-Dam-Westerhoff-96,Prindle_etal-12,Eckhardt_et_al-07}. The paradigmatic
model in this field is the Kuramoto-Sakaguchi model of globally coupled phase oscillators,
describing a transition to synchronization if
the attractive coupling is strong enough to overcome the natural spreading of oscillators
frequencies~\cite{Kuramoto-75,Sakaguchi-86}. The global coupling
typically appears in two setups. In one situation there are many links mutually connecting the oscillators in the population,
so that the all-to-all coupling is a suitable description. Such systems are widely considered in neuroscience, where neurons are
connected by an enormous number of synapses. In physical applications, in many cases the second
setup is relevant, where the global
coupling is due to a ``global mode''
which is fed by the units and acts back on them. For example, for Josephson junctions, electronic and spin-torque
oscillators, the global coupling is due to a
global current~\cite{Wiesenfeld-Colet-Strogatz-96,Temirbayev_etal-12,Grollier-Cros-Fert-06} which flows through the units in series; for lasers
the coupling is due to a global optical mode~\cite{Nixon_etal-13}; for metronomes,
pendulum clocks, and
for pedestrians on a bridge this
global mode is the oscillation mode of the support~\cite{Czolczynski_etal-13,Eckhardt_et_al-07}. Note that the global signal
can be rather complex and,
in particular, include several harmonics of the basic frequency~\cite{Daido-96}. For example, horizontal oscillations of the support lead to
a standard Kuramoto-Sakaguchi coupling of pendulum clocks via the first harmonics, while vertical oscillations produce
a second-harmonics coupling~\cite{Komarov-Pikovsky-13a,Komarov-Pikovsky-14}.
We nevertheless will dub this situation as one-mean-field (one-channel)
global coupling, as there is only one mediator for the global mode.

Global coupling can also differently force different oscillators, e.g., if the global
optical mode is oblique, an array of lasers will driven with different phase shifts. Another example is the coupling via
an acoustic/electronic receiver-emitter scheme~\cite{Vlasov-Macau-Pikovsky-14}, where the phase shifts experienced
by oscillators depend on the propagation time of
the global signal from emitter.

In this paper we generalize the Kuramoto coupling scheme and
study the dynamics of oscillators driven by \textit{two mean fields} via two different channels. A possible setup could be
electronic/acoustic oscillators, two mean fields of which are collected by two receivers that drive the oscillators via two emitters.
In optics, the two-channel coupling can be accomplished via splitting the global light mode beam and feeding  it back with
different phase shifts for the two parts. 
In life systems, in particular in coupled oscillating cells~\cite{Prindle_etal-12}, two global
coupling channels can be realized with two different messengers that carry signals from the cells into the mixing environment,
with possibly two different chemical mechanisms of influencing oscillations in the 
cells.  For example, in neuronal system, 
the signals between neurons can be transfered by both chemical transmitters 
and electrical coupling channels. So the neurons are actually coupled 
by two mean fields. Interaction through different channels is
also characteristic for physiological problems, where, e.g.,
cardiac and respiratory systems show coexisting
couplings \cite{Bartsch-etal-2012,Bartsch-Ivanov-2014,Bartsch-Liu-Ma-Ivanov-2014,Bartsch-Ivanov-NDES-2014}.
Another situation with two mean-field couplings, is related to the attempts to control synchrony in a population of coupled
oscillators~\cite{Rosenblum-Pikovsky-04b,Rosenblum-Pikovsky-04c}. Such an approach has been e.g. discussed in the context
of suppression of collective brain oscillations at Parkinson's
disease~\cite{Popovych-Hauptmann-Tass-05,Tukhlina-Rosenblum-Pikovsky-Kurths-07,Popovych-Tass-10}.
Here the two channels of coupling are the internal (uncontrollable) one, and the external coupling
due to an imposed feedback. Both couplings can be considered to a good approximation as global ones, acting on
the whole population of the involved neurons.

  Potentially, the most simple experimental realization of the two mean field coupling
 would be an extension of a recent experimental setup where Kuramoto-Sakaguchi coupling scheme with linear and nonlinear
 couplings have been experimentally verifyed~\cite{Temirbayev_etal-13}.  The scheme for the two-field coupling is presented in Fig. 1
 of the supplementary material. Instead of one common resistive load, like in the experiment~\cite{Temirbayev_etal-13},
 one could implement two such loads, and additionally modify the phase shifts of actions of the second mean field on the oscillators.
 This would exactly correspond to the particular model we study theoretically below.

We focus below on the simplest possible setup, where all oscillators that are described in the phase
approximation have the same natural frequency; furthermore, the coupling is assumed to involve
the first harmonics of the oscillations only, like
in the standard Kuramoto-Sakaguchi model. We  show that mostly nontrivial dynamics of this system is
observed if the two mean fields act differently,
i.e. one is mainly attractive while another one is repulsive. We find uniformly rotating (traveling wave)
solutions and study their stability
in the thermodynamic limit, where also the Ott-Antonsen approach can be applied \cite{Ott-Antonsen-08}. In the case
traveling waves are unstable, modulated traveling waves are observed. The findings are
supported by direct numerical simulations
of finite ensembles.

\section*{Results}
\section*{Model formulation}
In this paper we consider a simple model of a population of phase oscillators subject to a coupling through two
mean fields.  We start by formulating the problem in a rather general context, but will make several simplifications
to achieve a tractable but still nontrivial model. We assume that all the phase oscillators  have the same
natural frequency, and differ only by the way how they contribute to the mean fields and how they are forced by them.
Furthermore,
we assume that these differences are only in the phases of the coupling, not in the amplitudes. Physically, this
can be modeled by
 an ensemble of acoustic oscillators, sounds of which are collected by two microphones, and which are subject to the forcing
 emitted by two loudspeakers. If the positions of the oscillators are different, they contribute to the mean fields with different
 phase shifts and get signals that are also differently phase shifted. On the other hand, attenuation of signals can be neglected,
 thus only the phase relations are important.

 We denote the phases of the oscillators $\phi_k$, $k=1,\ldots,N$, and define two complex mean fields $Y^{(1,2)}$ according to
 \begin{equation}
 Y^{(1,2)}=\frac{1}{N}\sum_{j=1}^N e^{i\phi_j+i\gamma_j^{(1,2)}}
 \label{eq:mf}
 \end{equation}
 where $N$ is the number of oscillators in the system, and $\gamma_j^{(1,2)}$ are the phase shifts with which the oscillators contribute to the mean fields. The dynamics
 of the phases, driven by these fields, is given by equations (written in the reference frame rotating with the common frequency)
 \begin{equation}
 \dot\phi_k=\varepsilon_1 \text{Im}(Y^{(1)}e^{-i\phi_k-i\delta^{(1)}_k})+
 \varepsilon_2\text{Im}(Y^{(2)}e^{-i\phi_k-i\delta^{(2)}_k})
 \label{eq:bm1}
 \end{equation}
 where $\varepsilon_{1,2}$ are the coupling constants of the two fields, and $\delta^{(1,2)}_k$ are the phase shifts with which the
 fields act on oscillators.

 Generally, the model above would be fully defined if the joint distribution density of the phase shifts
 $W(\gamma^{(1,2)},\delta^{(1,2)})$ is given. To simplify, we assume that the parameters $\gamma^{(1,2)},\delta^{(1,2)}$ are
 independent on each other in the populations. Due to this, as one can see from \eqref{eq:bm1}, the dynamics of $\phi_k$ does not
 depend on the phase shifts $\gamma^{(1,2)}$ and one can simplify the expressions for the mean fields \eqref{eq:mf} as
 \begin{equation}
 Y^{(1,2)}=\langle e^{i\phi}\rangle\langle e^{i\gamma^{(1,2)}}\rangle=Z w^{(1,2)}\exp[i\mu^{(1,2)}]
 \label{eq:mf2}
  \end{equation}
 where we introduced the usual Kuramoto complex mean field and two complex constants
 characterizing the distributions of the phase shifts   $\gamma^{(1,2)}$ :
 \begin{equation}
 Z=\langle e^{i\phi}\rangle=\frac{1}{N}\sum_j e^{i\phi_j}\;,\qquad  w^{(1,2)}\exp[i\mu^{(1,2)}]=\langle e^{i\gamma^{(1,2)}}\rangle
 \label{eq:mf3}
  \end{equation}

 The dynamics of the phases~\eqref{eq:bm1} remains depending on the distributions of $\delta^{(1,2)}_k$. Below we
 consider a minimal nontrivial case, when only one set of the phase shifts $\delta^{(1,2)}_k$ has a nontrivial distribution
 (say, the second one), while
 another set consists of equal shifts  $\delta^{(1)}$. Then we can rewrite Eq.~\eqref{eq:bm1} as
 \begin{equation}
 \dot\phi_k=\varepsilon_1 w^{(1)}\text{Im}(Z e^{-i\phi_k-i\delta^{(1)}+i\mu^{(1)}})+
 \varepsilon_2 w^{(2)}\text{Im}(Z e^{-i\phi_k-i\delta^{(2)}_k+i\mu^{(2)}})
 \label{eq:bm2}
 \end{equation}
We renormalize time $t \varepsilon_2 w^{(2)}\to t$ and obtain finally the basic model that we will study in this paper:
\begin{equation}
 \dot\phi_k=\lambda \text{Im}(Z e^{-i\phi_k+i \Theta_1})+
 \text{Im}(Z e^{-i\phi_k+i\alpha_k+i\Theta_2}),\quad Z=Re^{i\Psi}=\langle e^{i\phi}\rangle
 \label{eq:bm3}
 \end{equation}
 where we introduced a real parameter $\lambda=\frac{\varepsilon_1 w^{(1)}}{ \varepsilon_2 w^{(2)}}$ describing the relative
 strengths of couplings of the two mean fields; $\Theta_1=\mu^{(1)}-\delta^{(1)}$ is the effective constant phase shift for the
 first mean field; $\Theta_2$ is the phase at which the distribution of the phase shift of the second field
 $\mu^{(2)}-\delta^{(2)}_k$ has a maximum; and finally the parameter
 $\alpha_k=\mu^{(2)}-\delta^{(2)}_k-\Theta_2$ describes the deviation of the phase shift from this most probable one.
 Below we use the von Mises distribution of phase shifts $\alpha$
 \begin{equation}
 g(\alpha)=\frac{\exp[\Delta\cos\alpha]}{2\pi I_0(\Delta)}
 \label{eq:vmd}
 \end{equation}
 characterized by the parameter $\Delta$: $\Delta=0$ corresponds to a uniform distribution of the phase
 shifts, while in the limit $\Delta\to\infty$ one gets a delta-function.

 It is instructive to rewrite Eq.~\eqref{eq:bm3} in the real form:
 \begin{equation}
 \dot\phi_k=\lambda R\sin(\Psi-\phi_k+\Theta_1)+R\sin(\Psi-\phi_k+\Theta_2+\alpha_k),\qquad Re^{i\Psi}=\langle e^{i\phi}\rangle
 \label{eq:bm4}
 \end{equation}
The physical interpretation of this two-mean-field-coupling model~\eqref{eq:bm4}
is as follows: each of the couplings
is of Sakaguchi-Kuramoto type, i.e. it contains the sin term only. The two couplings have
different phase shifts,
for the first coupling it is fixed to $\Theta_1$, for the second coupling the phase shift is
different for different oscillators, but
is concentrated around $\Theta_2$. Parameter $\lambda$ defines the relative weight of the
two couplings. The main parameters
of the problem are $\lambda,\Theta_1,\Theta_2$, and $\Delta$. One can expect mostly
nontrivial effects if
the two couplings act in opposite directions: one tries to synchronize the oscillators, while
another one is repulsive and tends
to desynchronize them. Below we will keep all the parameters in the theoretical considerations, but
in numerical
examples we will mainly set either
$\Theta_1=0$, that corresponds to a purely attractive first coupling, or $\Theta_2=0$, that
corresponds to a purely attractive second coupling.

\section*{Stability of incoherent state}
\begin{figure}
\centering
\includegraphics[width=\columnwidth]{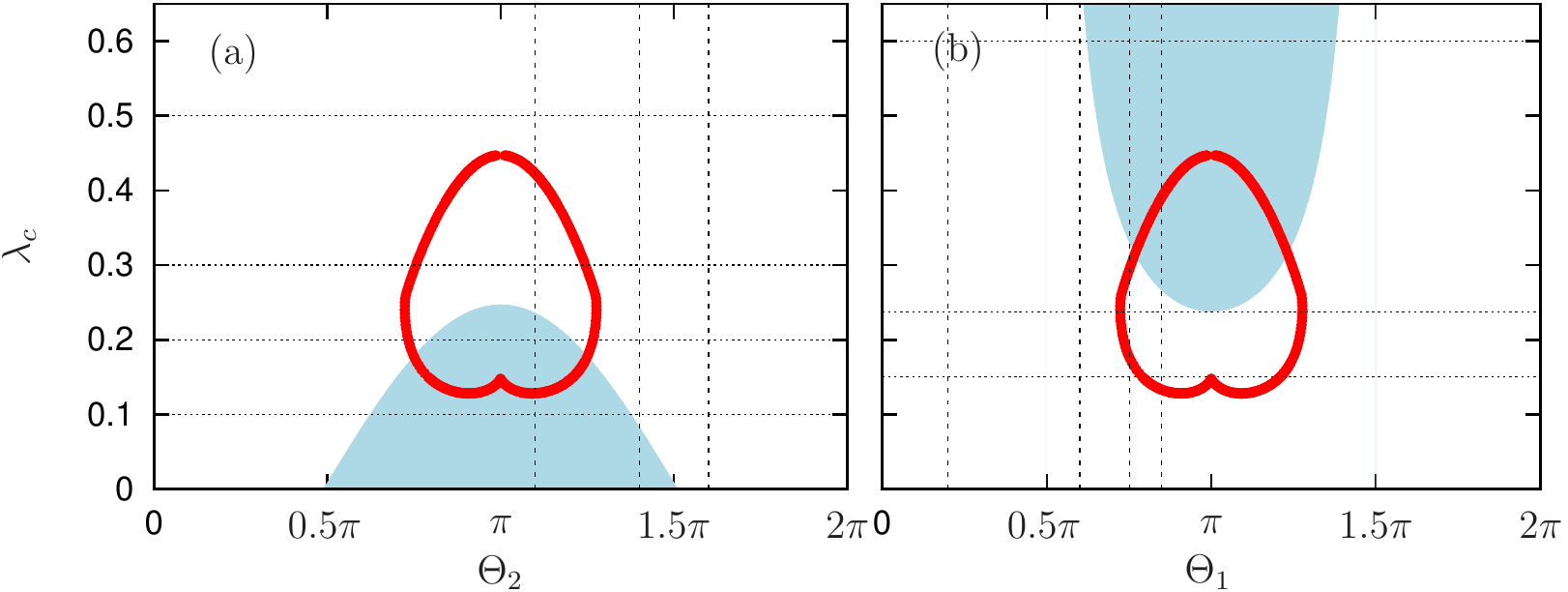}
\caption{\textbf{Stability of the incoherent state and the end points of anomalous branches.} Blue region: stability of the incoherent state according to formula \eqref{eq:stbo}.
Red lines: points on plane of parameters $(\Theta_{1,2},\lambda)$, where $R=0$.
These are the points (calculated according to formula \eqref{eq:aav1})
where the ``anomalous'' branch has vanishing order parameter.
Panel (a): case $\Theta_1=0$; panel (b): case $\Theta_2=0$.
Dotted lines show values of $\lambda$ for which the diagrams in
Figs.~\ref{fig:scsol} and \ref{fig:scsol2} are plotted. Dashed lines show values of $\Theta_1,\Theta_2$
for which the diagrams in  Figs.~\ref{fig:scsol1} and \ref{fig:scsol3}
are plotted.}
\label{fig:linst}
\end{figure}

First, we consider stability properties of the completely incoherent state where the phases of
all the oscillators are uniformely distributed and the mean field vanishes.
Analytical expression for the growth rate of potentially unstable perturbations, derived in the section Methods,
reads
\begin{equation}
\text{Re}(\gamma)=\frac{\lambda \cos \Theta_1}{4}+\frac{I_1(\Delta)\cos \Theta_2}{4I_0(\Delta)}
\label{eq:stbo1}
\end{equation}
It determins critical coupling parameter $\lambda_c$ at which the incoherent state becomes unstable (Re$(\gamma)>0$).
For example, for $\Theta_1=0$ (attractive first coupling) we
have $\lambda_c=-\cos \Theta_2\frac{I_1(\Delta)}{I_0(\Delta)}$, this boundary is depicted
in Fig.~\ref{fig:linst}(a). One can see that stability of the incoherent state is only possible
if the second coupling is repulsive (i.e. $\Theta_2$ is close to $\pi$) and
relatively strong ($\lambda$ is small). In panel (b) we show the
case when $\Theta_2=0$ is fixed, here the
stability boundary of the incoherent state is
given by $\lambda_c=-\frac{I_1(\Delta)}{I_0(\Delta)\cos\Theta_1}$.
In this figure there is another nontrivial line to be discussed below in the next section.

\section*{Traveling wave solutions and their stability}
Here we discuss nontrivial regimes of partial synchrony in the model. We use the self-consistent
approach to find the
solutions; reformulate the system in terms of Ott-Antonsen (OA) equations; and then determine stability of
the solutions by calculating their stability spectra. The details of these methods are described in the Methods
Section, here we summarise the results.

The traveling wave synchronous solution have the mean field with a constant amplitude $R$ and a uniformly rotating
phase $\Psi=\Omega t+\Psi_0$. The distribution
of the phases is stationary in the reference frame rotating with $\Omega$. Driven by the two mean fields, some
oscillators are locked (i.e. they rotate with the same frequency $\Omega$), while others are not  locked and rotate
(although non-uniformly) with some other frequencies.
The main parameters we use to characterize the state of the system, are the amplitude $R$ and the locking
parameter $P_l$ (cf. Eq.~\eqref{eq:plock}), determining which portion of oscillators in the population is locked ($P_l=1$
means all are locked; $P_l=0$ means no one is locked). For determination of stability, we use a reformulation of the
dynamics in terms of the Ott-Antonsen integro-differential system of equations (see description in Methods Section, in
particular Eq.~\eqref{eq:oac}); the latter equation is linearized and discretized to find the spectrum of eigenvalues.

\begin{figure}[!ht]
\centering
\includegraphics[width=\columnwidth]{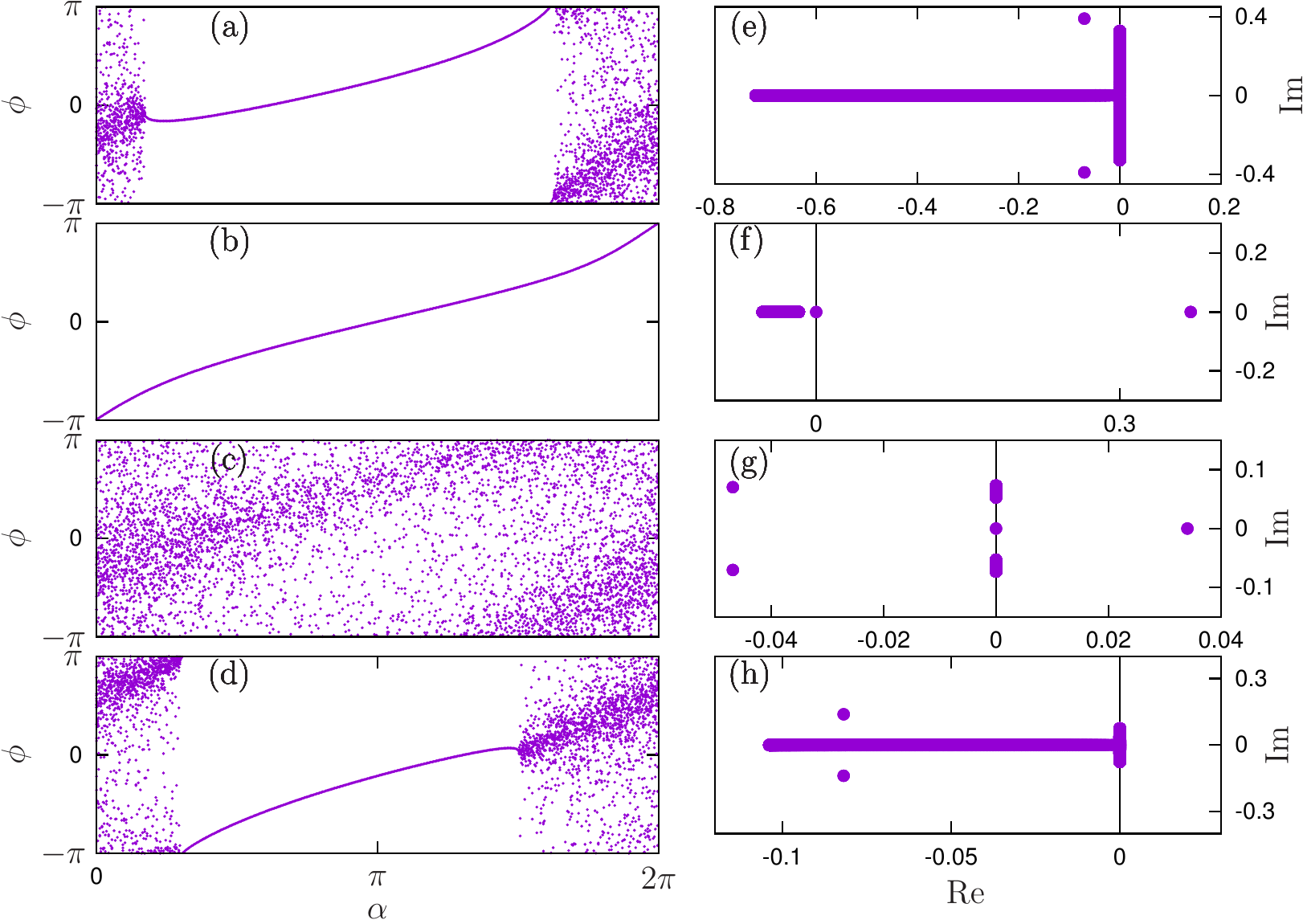}
\caption{\textbf{Synchronous states and their stability.}
Phase profiles (left panels) and their stability spectra (right panels) for the constructed
synchronization states.
Panels (a,b,e,f) show the case of attractive first coupling.
The parameters here are $\Theta_1=0$, $\Theta_2=-0.9\pi$, and $\lambda=0.45$, with
(a): $R=0.295$ and the locking parameter $P_l=0.61$ (normal branch);  and
(b): $R=0.02$ and $P_l=1$ (anomalous branch). (e,f): correspondent stability spectra of these states. Vertical line shows
the imaginary axis,
for better identification of instability.
Panels (c,d,g,h) show the case of attractive second coupling with
parameters $\Theta_1=0$, $\Theta_2=-0.9\pi$ and $\lambda=0.2$, with (c):
$R=0.028$ and $P_l=0$ (normal branch); and (d)
$R=0.07$ and $P_l=0.45$ (anomalous branch).
}
\label{fig:fields1}
\end{figure}


In Fig.~\ref{fig:fields1} we present some typical stationary  synchronous states together
with their stability
spectra. In these figures we visualize the stationary density
via a dispersion of a finite set of points, for better visibility. One can easily distinguish domains of locked (all points
representing oscillators collapse to a line) and rotating (scattered) oscillators.  We must 
emphasize that the partial phase locked state here is not a chimera state, as the oscillators react to the fields with 
different phase shifts, and therefore there is no structural symmetry in the system which has to be broken in the chimera phenomenon. Panel (a), (c), (e) and (g) show the profile of normal branches which can also be found in one mean field coupling cases, while the anomalous branches shown in panel (b), (d), (f) and (h), seldom appear in the traditional one mean field case. One remarkable thing is that, the ratio of phase locked oscillators in (b) is larger than in (a), but the order parameter in (b) is smaller than in (a). This is also different from the usual cases.

One can see from Fig.~\ref{fig:fields1} that the spectrum generally
consists of  a continuous part and several discrete  eigenvalues. The continuous
part is related to existence of the branch of the locked oscillators (a set of purely real eigenvalues)
and to the unlocked oscillators (a set of
purely imaginary eigenvalues). In the case both branches are present (panels (a,d) of Fig.~\ref{fig:fields1}), one has a
characteristic T-shaped continuous (essential)
spectrum, as argued in Refs.~\cite{Wolfrum_etal-11,Omelchenko-13} for the chimera states,
where also locked and unlocked oscillators are present.
The state in panel (b) of  Fig.~\ref{fig:fields1} is fully locked, here only the real continuous
spectrum is present.  In the case (c) of  Fig.~\ref{fig:fields1},  where all the oscillators are unlocked,
the purely imaginary
continuous spectrum is observed.

As argued in~Refs.~\cite{Wolfrum_etal-11,Omelchenko-13}, stability is determined by the
discrete spectrum, which is clearly seen for all cases: panels  (a,d)
depict
stable solutions, while panels (b,c) depict unstable ones.
In some cases (not illustrated here) it is difficult to distinguish
the discrete spectrum from the continuous one, as the discrete eigenvalues have very small
real part and are ``smeared'' (i.e. depend
significantly on the offset parameter $\alpha_0$). Here refining the resolution (i.e. increasing
the number of discretization points $L$)
helps, but the computation time increases rapidly.

\section*{Normal, anomalous, and hybrid states}
Basing on the approach described above, we have determined the
uniformly rotating states for different values
of parameters, and characterized their stability. Here below we describe the main types of the solutions.

\begin{figure}[!ht]
\centering
\includegraphics[width=0.7\columnwidth]{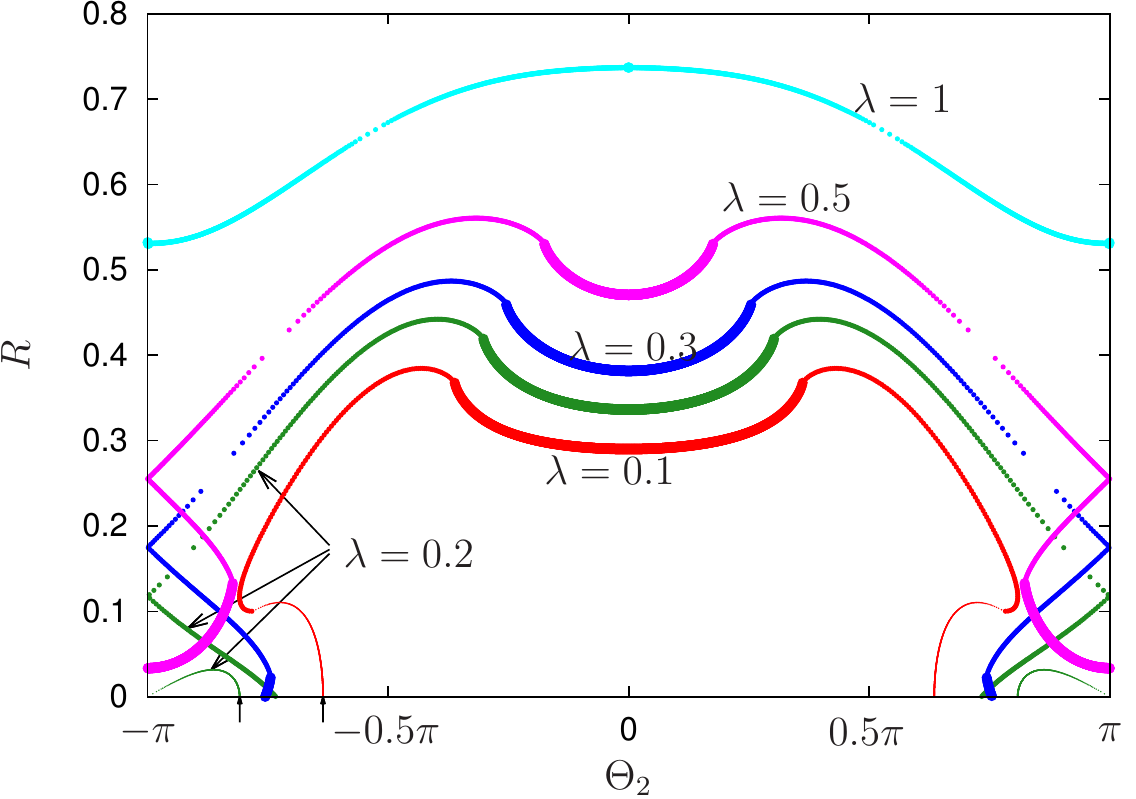}
\caption{
\textbf{Synchronous states for attractive first coupling.}
Order parameter $R$ for $\Theta_1=0$ and $\Delta=0.5$, as functions of $\Theta_2$
for several values of $\lambda$. Large markers: states with all $P_l=1$ (all oscillators locked);
medium markers: states that are partially locked $0<P_l<1$; small markers: all oscillators are
unlocked $P_l=0$. Arrows at the $x$-axis
show the linear stability boundaries for $\lambda=0.1$ and for $\lambda=0.2$, according to Eq.~\eqref{eq:stbo}.}
\label{fig:scsol}
\end{figure}

\begin{figure}
\centering
\includegraphics[width=0.7\columnwidth]{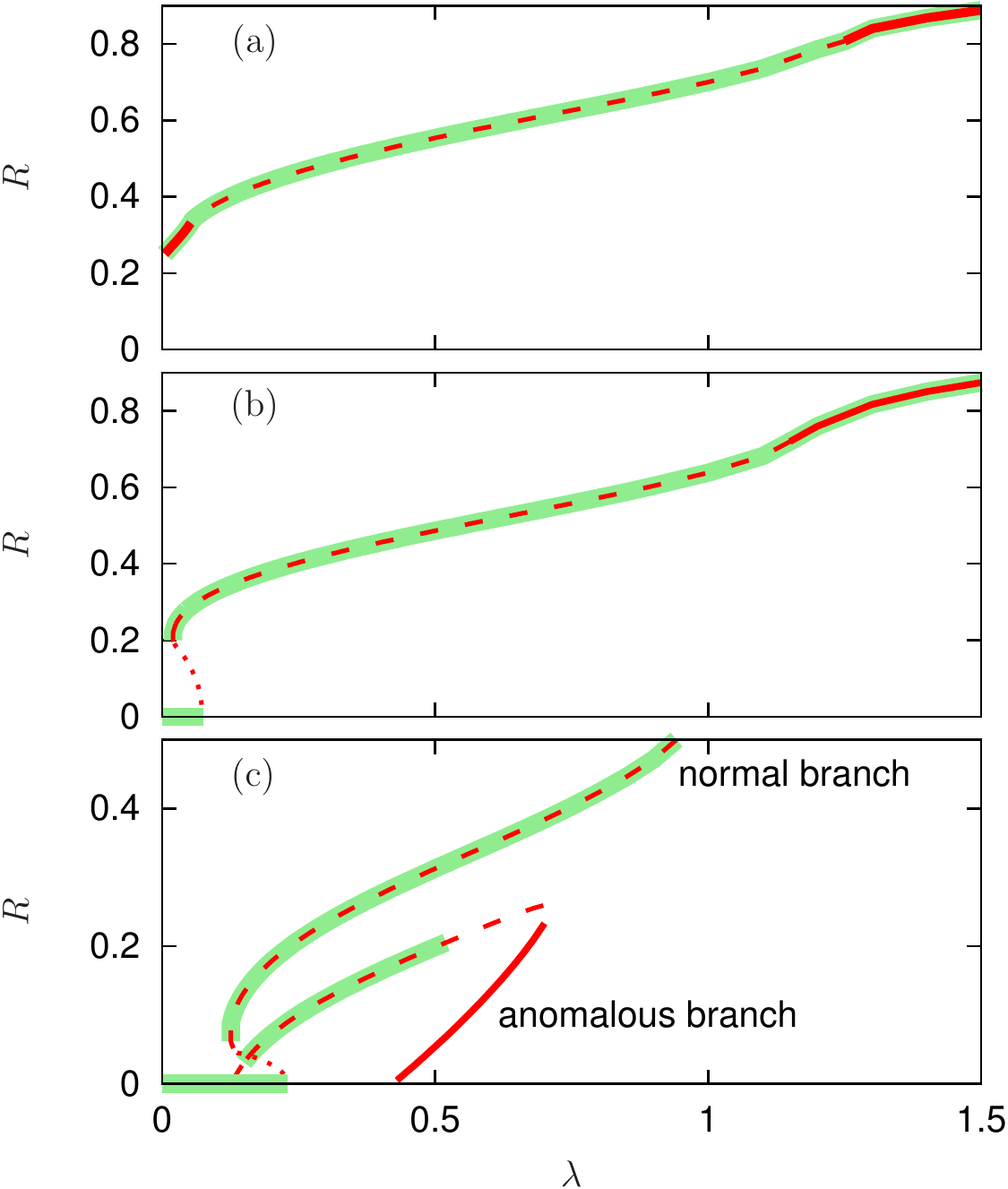}
\caption{\textbf{Normal and anomalous branches for attractive first coupling.}
The global order parameter of traveling wave states as functions of $\lambda$.
The parameters are  $\Theta_1=0$, $\Delta=0.5$,  and the values of $\Theta_2$ are (a): $\Theta_2=-0.4\pi$; (b): $\Theta_2=-0.6\pi$;
(c): $\Theta_2=-0.9\pi$. Bold lines: states with all $P_l=1$ (all oscillators locked);
dashed lines: states that are partially locked $0<P_l<1$; dotted lines:
all oscillators are unlocked $P_l=0$.  Stable traveling wave and incoherent
solutions are highlighted with bold green line in background.
}
\label{fig:scsol1}
\end{figure}

\subsection{Attractive first coupling}

In this subsection we present the results for the case $\Theta_1=0$, i.e. the first (non-distributed) coupling is purely attractive.
We present the dependencies of the order parameter $R$ on $\Theta_2$, obtained using the self-consistent method
above, in Fig.~\ref{fig:scsol}.   To fix $\Theta_1=0$,
in relations~\eqref{eq:rf} we varied parameter
$\beta$ in the range $0\leq\beta<2\pi$ and found all the points with $Q=\Theta_1=0$.
These points yield a parametric representation
of the order parameter $R$ and of $\Theta_2$ as functions of the remaining auxiliary parameter $a$. Additionally, we calculated
the portion of locked oscillators according to \eqref{eq:plock} and coded it in  Fig.~\ref{fig:scsol}
with the size of the markers into three types: all
oscillators locked ($P_l=1$), all oscillators unlocked ($P_l=0$), and partial locking ($0<P_l<1$).
First, one can notice the symmetry of the diagram Fig.~\ref{fig:scsol} $\Theta_2\to -\Theta_2$,
this is the consequence of the selected
value $\Theta_1=0$. Therefore, for the analysis of the dependence of the order parameter $R$ on $\lambda$, presented in
 Fig.~\ref{fig:scsol1}, we choose only negative values
of the phase shift $\Theta_2$. In Fig.~\ref{fig:scsol1} we also use the style of the lines
(solid, dashed, dotted) to distinguish fully locked, partially locked, and fully unlocked states.

The situation is quite simple for $\Theta_2\approx 0$: here both mean fields act attractively and the dependence on the
parameter $\lambda$ is monotonous (see the curve for $\Theta_2=-0.4\pi$ in Fig.~\ref{fig:scsol1}(a)). Remarkably, here
solutions with small value of $\lambda$, where the second mean field dominates, are fully locked, while at medium values
of $\lambda$, where both fields compete, only partial locking is observed.
At large  $\lambda$ the attractive first coupling dominates again, and here $P_l=1$.

For phase shifts $\Theta_2$ that are
closer to $-\pi$, instead of a monotonous dependence of the order parameter $R$ on $\lambda$, we obtain a
dependence characteristic for
a subcritical transition of the first-order type, see the curve for $\Theta_2=-0.6\pi$
in Fig.~\ref{fig:scsol1}(b).
Stability consideration reveals that in the region where two nontrivial solutions exist,
solutions with lower
value of $R$ are unstable and those with higher value of $R$ are stable.  Thus, we have a
typical diagram for a hysteretic
transition of the first order: there is a range of parameter $\lambda$ where the incoherent
and coherent solutions are both stable,
while ``in between'' there is an unstable solution. When this unstable solution meets the
incoherent one (see arrows below the $\Theta_2$-axis
in Fig.~ \ref{fig:scsol}) the latter loses its stability.
It is worth mentioning that the large part of the unstable coherent solution is fully unlocked,
although a part
of it can be partially locked (e.g., this happens for $\Theta_2=-0.9\pi$, see
Fig.~ \ref{fig:scsol1}(c)).

The situation for $\Theta_2$ even closer to $-\pi$,
illustrated by the results
for $\Theta_2=-0.9\pi$ in Fig.~\ref{fig:scsol1}(c), is rather unusual and
deserves a more detailed
description. First, one can see here the branch which is very much similar to the solution for
$\Theta_2=-0.6\pi$, only it is shifted to larger values of parameter $\lambda$.
Here the stability
properties are similar to that at $\Theta_2=-0.6\pi$: solutions with $dR/d\lambda>0$
are stable while
the state with $dR/d\lambda<0$ is unstable. The latter unstable part of this branch appears
exactly at the linear stability border point. We call this branch ``normal branch'', as it is very similar
to the situation observed at a usual subcritical transition.

One can see that additionally there exists another branch of solutions with
relatively small values of $R$.  Noteworthy, the values of $\lambda$ at which
this branch emerges from $R=0$, have
nothing to do with the change of linear stability of the incoherent state. Therefore we call this
branch ``anomalous''. The anomalous solutions can also be seen in
Fig.~\ref{fig:scsol}, they do not end
at the points (marked by arrows) where linear stability of the incoherent state changes.

To clarify the behavior of the solutions in the limit $R\to 0$, we need to take this limit for
the solutions constructed by Self-consistent approach presented in the Methods Section. There are two possibilities to have vanishing
order parameter $R$: one is related to vanishing of $r$ in the equation for the phases
\eqref{eq:psieq}, and another is related to the case $\Omega=0$. The former, ``normal''
case corresponds
to a uniform distribution of the phases for all $\alpha$ in the limit $R\to0$, i.e. to the incoherent state. The values
of parameters at which this happens are exactly those at which the linear stability of the incoherent
state changes. Thus the ``normal'' branch bifurcates from the non-coherent state.

In contradistinction,
the states with vanishing $R$ that correspond to $\Omega=0$, are not close to the incoherent state
with a uniform distribution of phases. Moreover, as Fig.~\ref{fig:scsol1}(c) shows, one part of
the ``anomalous'' branch consists of states with $P_l=1$, i.e. all the oscillators are locked.
 Therefore we cannot describe the appearance of the ``anomalous'' branch as a bifurcation from the incoherent solution.

 Stability analysis of the anomalous branch shows that the solutions with $P_l=1$ are
unstable, while another part with $0<P_l<1$ consists
of unstable and stable
solutions.
The change of stability happens without appearance of new
traveling wave solutions, although we cannot exclude that some other solutions (e.g.,
modulated traveling waves) may appear at these points.

\subsection{Attractive second coupling}

\begin{figure}
\centering
\includegraphics[width=0.7\columnwidth]{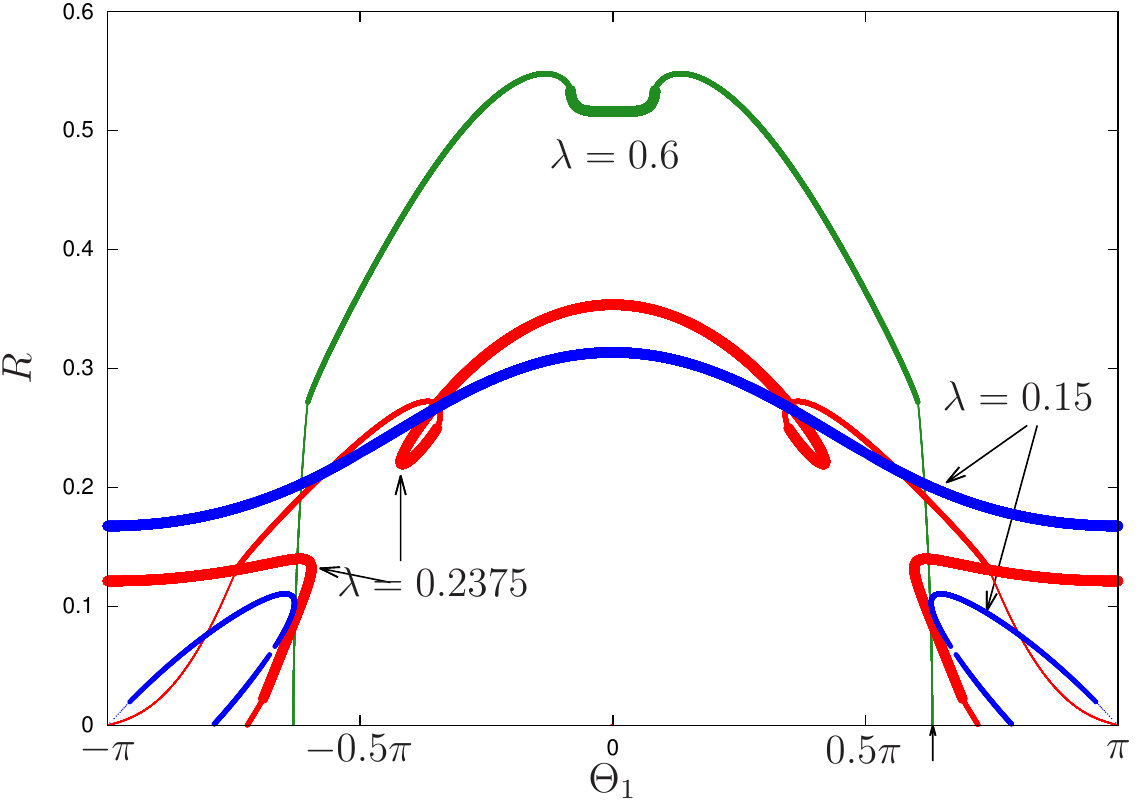}
\caption{\textbf{Synchronous states for attractive second coupling.}
The order parameter $R$ for $\Theta_2=0$ and $\Delta=0.5$, as functions of $\Theta_1$
for several values of $\lambda$. Large markers: states with all $P_l=1$ (all oscillators locked);
medium markers: states that are partially locked $0<P_l<1$; small markers: all oscillators are
unlocked $P_l=0$. Arrows at the $x$-axis
show the linear stability boundaries for $\lambda=0.6$ according to Eq.~\eqref{eq:stbo}.}
\label{fig:scsol2}
\end{figure}

\begin{figure}
\centering
\includegraphics[width=0.9\columnwidth]{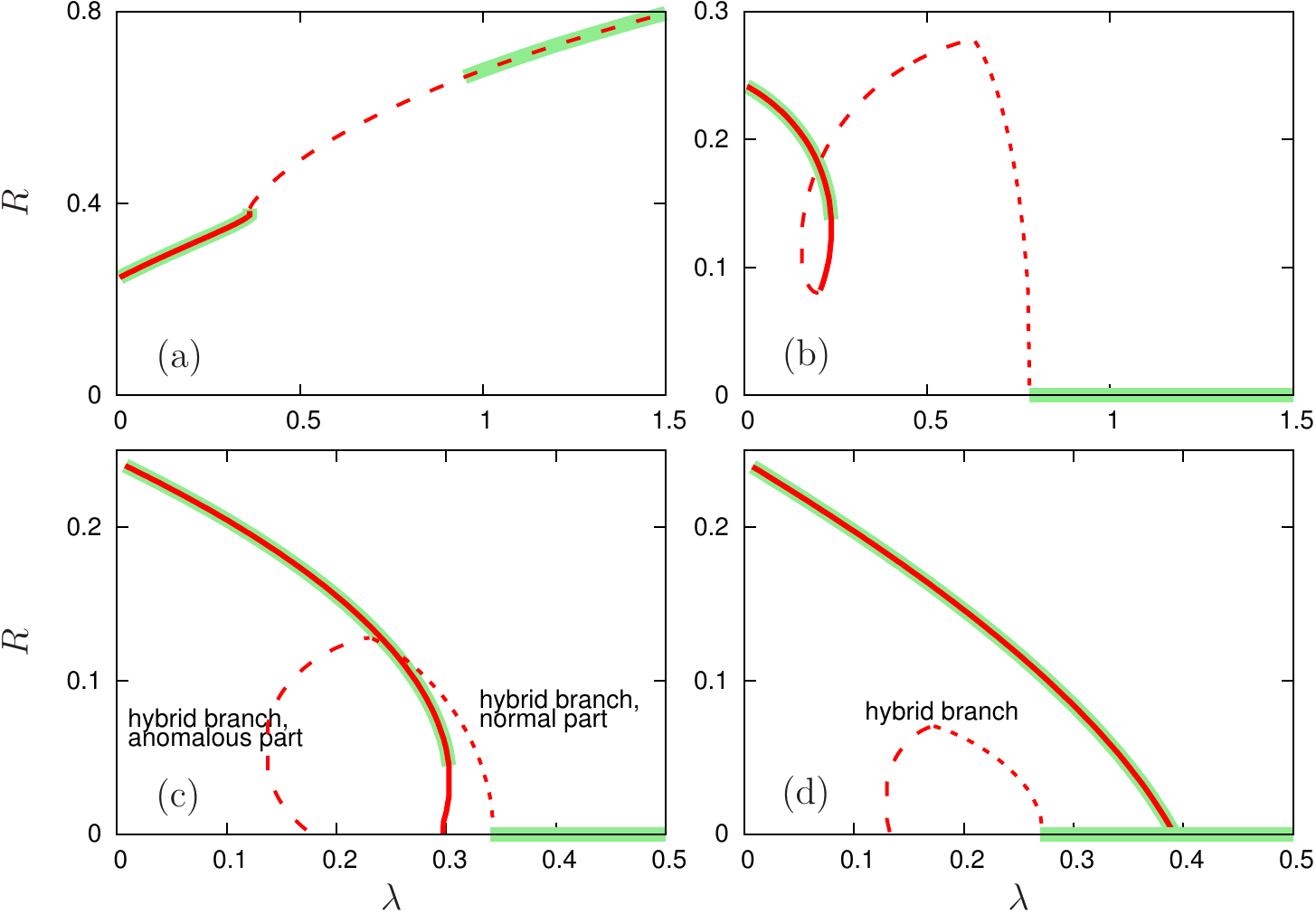}
\caption{\textbf{Normal, anomalous, and hybrid branches for attractive second coupling.}
The global order parameter of traveling wave states as functions of $\lambda$. The parameters are  $\Theta_2=0$ and $\Delta=0.5$, and the values of $\Theta_1$ are (a): $\Theta_1=0.2\pi$; (b): $\Theta_1=0.6\pi$;
(c): $\Theta_1=0.75\pi$; (d): $\Theta_1=0.85\pi$. Bold lines: states with all $P_l=1$ (all oscillators locked);
dashed lines: states that are partially locked $0<P_l<1$; dotted lines:
all oscillators are unlocked $P_l=0$.  Stable traveling wave and incoherent
solutions are highlighted with bold green line in background.
}
\label{fig:scsol3}
\end{figure}

Here we discuss a situation where the phase shift of the second coupling is fixed $\Theta_2=0$,
while the phase shift of
the first coupling $\Theta_1$ varies.
Figures~\ref{fig:scsol2},\ref{fig:scsol3} are similar to the Figs.~\ref{fig:scsol},\ref{fig:scsol1} in the
previous subsection. Fig.~\ref{fig:scsol2} shows nontrivial states when the first and the second couplings
act in different directions, i.e. when $\Theta_1$ is close to $\pi$.
Let us focus on the dependence of the order parameter $R$ on the coupling constant $\lambda$
in Fig.~\ref{fig:scsol3}. For small values of $\Theta_1$, the incoherent state is unstable for all
$\lambda$, and a traveling wave solution exists - however it is unstable for intermediate values of
$\lambda$, see  Fig.~\ref{fig:scsol3}(a).  For larger values of $\Theta_1>\pi/2$,
where the first coupling is repulsive, the incoherent state becomes stable at large $\lambda$
(see Fig.~\ref{fig:scsol3}(b), here $\Theta_1=0.6\pi$).
When $\lambda$  decreases, at the instability border the normal branch of solutions softly appears
with small values of $R$.
With decreasing of $\lambda$ this branch makes a loop, and only for small $\lambda$ a part
of this branch becomes stable. For larger values of $\Theta_1$, the loop becomes larger
and eventually crosses
the axis $R=0$ at two points (which correspond to the situations where anomalous solutions have
vanishing order parameter, these points are marked red
in Fig.~\ref{fig:linst}(b)); see also panels (c,d) in  Fig.~\ref{fig:scsol3}. Remarkably, as a result the branch
which starts as a normal branch, now ends as an anomalous one, thus it can be termed ``hybrid
branch''. Another anomalous branch that consists of solutions with $P_l=1$, ends at $\lambda=0$.
The anomalous and the hybrid branches exist also for $\Theta_1=0.85\pi$, see
Fig.~\ref{fig:scsol3}(d).

Noteworthy, panels (a,b,c) of Fig.~\ref{fig:scsol3} show stability gap: there are intermediate values of
the coupling constant $\lambda$ at which there are no stable traveling wave solutions. Here
numerical simulations, presented in the next sections, demonstrate time-periodic regimes (modulated
traveling waves). In Fig.~\ref{fig:scsol3}(d) there is a bistable region where
both the anomalous and the incoherent solutions are stable, here a hysteretic transition is observed.

\section*{Simulation of the Ott-Antonsen system}
\label{sec:oasim}

\begin{figure}[!ht]
\centering
\includegraphics[width=0.5\columnwidth]{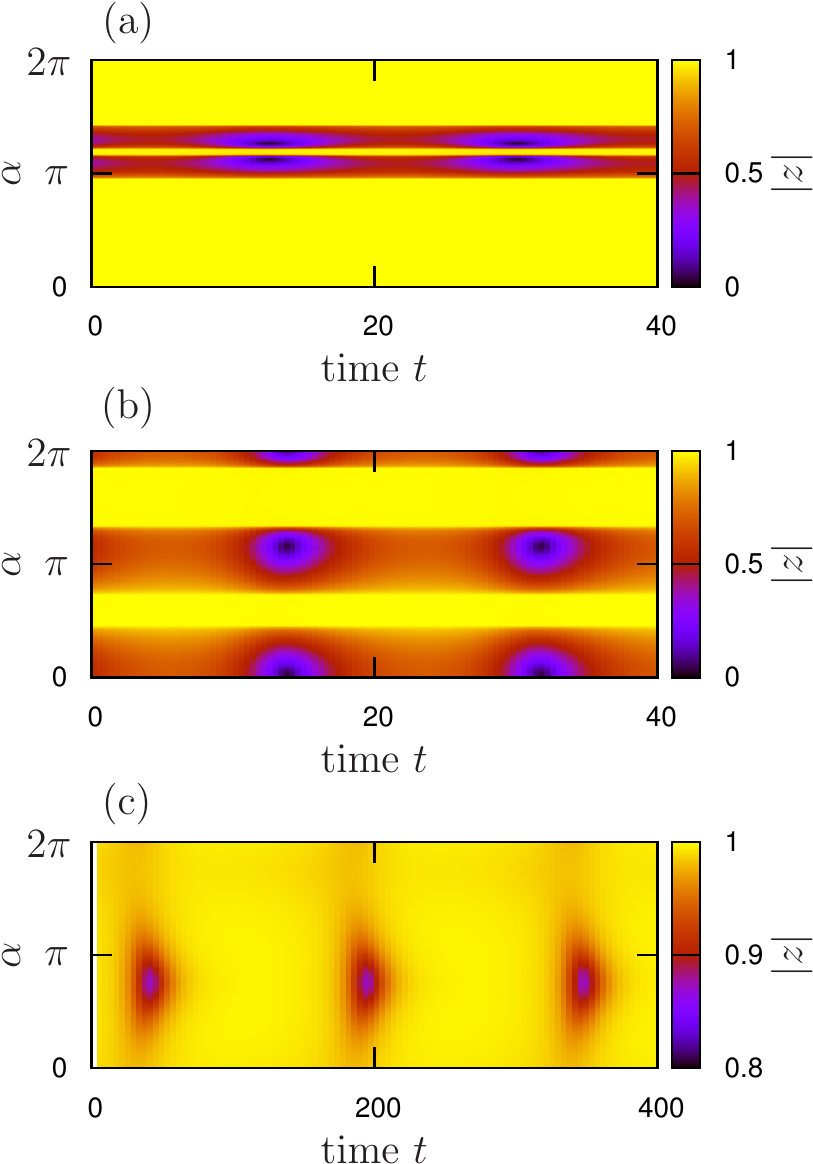}
\caption{\textbf{Modulated solutions in simulations of Ott-Antonsen equations.}
Periodic in time evolution of the  local order parameter $|z(\alpha,t)|$ for  situations
where stationary traveling waves are unstable.
Panel (a): $\Theta_1=0.2\pi$, $\Theta_2=0$ and $\lambda=0.7$.  Panel (b): $\Theta_1=0.6\pi$, $\Theta_2=0$ and $\lambda=0.5$.
Panel (c): $\Theta_1=0.75\pi$, $\Theta_2=0$ and $\lambda=0.315$.
Simulations started from random initial conditions, the transient stage is omitted. The same parameter values are used in Fig.~\ref{fig:snapshot}.
}
\label{fig:per}
\end{figure}

In this and the next section we report on the numerical tests of the found solutions.
The first test deals with the system in the thermodynamic limit. Here, according to the Ott-Antonsen
ansatz~\cite{Ott-Antonsen-08}, one can write for each $\alpha$ a closed equation for the local order
parameter $\langle e^{i\psi_\alpha}\rangle$ and thus to represent the whole system as an integro-differential equation (equation ~\eqref{eq:oac} in the Method Section).
The simulation is accomplished via discretization of the integral and by solving the resulting
finite-dimensional system with the Runge-Kutta method. In the cases, when at least one stable traveling wave solution
is present, direct numerical simulations confirm the stability results above.
Thus we present only the nontrivial cases where no stable traveling wave solution exists (see Fig.~\ref{fig:scsol3} above).
In Fig.~\ref{fig:per}  we show the patterns appearing in these situations. The dynamics of the order parameter $|z(\alpha,t)|$
is periodic in time. Together with rotation of the phase this gives quasiperiodic dynamics (cf. Fig.~\ref{fig:snapshot} below where the same
patterns are checked in finite-size ensemble simulations), i.e. a modulated traveling wave. Remarkably, patterns in panels (a,b) of
Fig.~\ref{fig:per}   show two locked and two unlocked regions. Modulated wave in panel (c) of
Fig.~\ref{fig:per}   can be characterized as a ``blinking locked state'',
as here for large time intervals the local order parameter is close to one everywhere, and only for
relatively short time intervals,
weakly unlocked regions, where this order parameter is reduced to $|z|\approx 0.9$, are observed.

\section*{Direct numerical simulations of oscillator population}

\begin{figure}[!hbt]
\includegraphics[width=0.5\columnwidth]{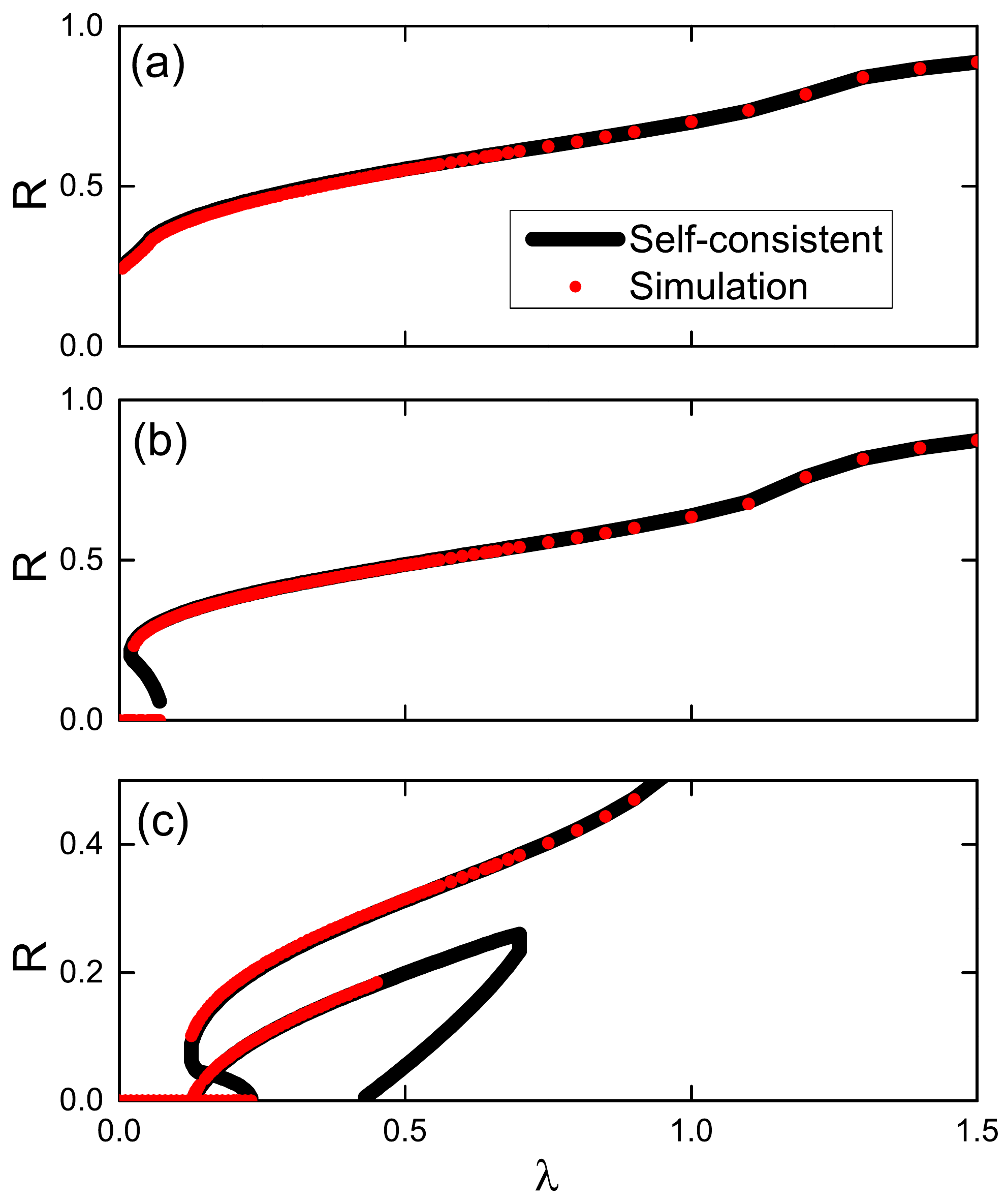}
\caption{\textbf{Direct numerical simulations for attractive first coupling.}
The global order parameter of traveling wave states as functions of $\lambda$.
The parameters are $\Theta_1=0$, $\Delta=0.5$, and the values of $\Theta_2$ are (a): $\Theta_2=-0.4\pi$; (b): $\Theta_2=-0.6\pi$; (c): $\Theta_2=-0.9\pi$. Black lines represent the self-consistent solution and red circles represent the direct simulation results.}
\label{fig:R-lambda1}
\end{figure}

In this section, we present results for numerical simulations of a finite size oscillator system, Eq.~\eqref{eq:bm4}.
First, we present results for a large system: N=10000. Fig. 2 in the Supplementary Information shows both the results from the
self-consistent solution and direct simulation result with the parameter $\Theta_1=0$ (attractive first mean field).
The results confirm our solutions in the thermodynamic limit, presented in Fig \ref{fig:scsol}.

Fig. \ref{fig:R-lambda1} compares the self-consistent solutions
and the simulation results for the traveling wave states shown in \ref{fig:scsol1} above. The parameters are $\Theta_1=0$, $\Delta=0.5$ and $\Theta_2=$ (a): $-0.4\pi$; (b): $-0.6\pi$;
(c): $-0.9\pi$. The simulation results show the normal and the anomalous solutions in Fig. \ref{fig:scsol1}, and also confirm our previous results.

\begin{figure}[!h]
\includegraphics[width=0.7\columnwidth]{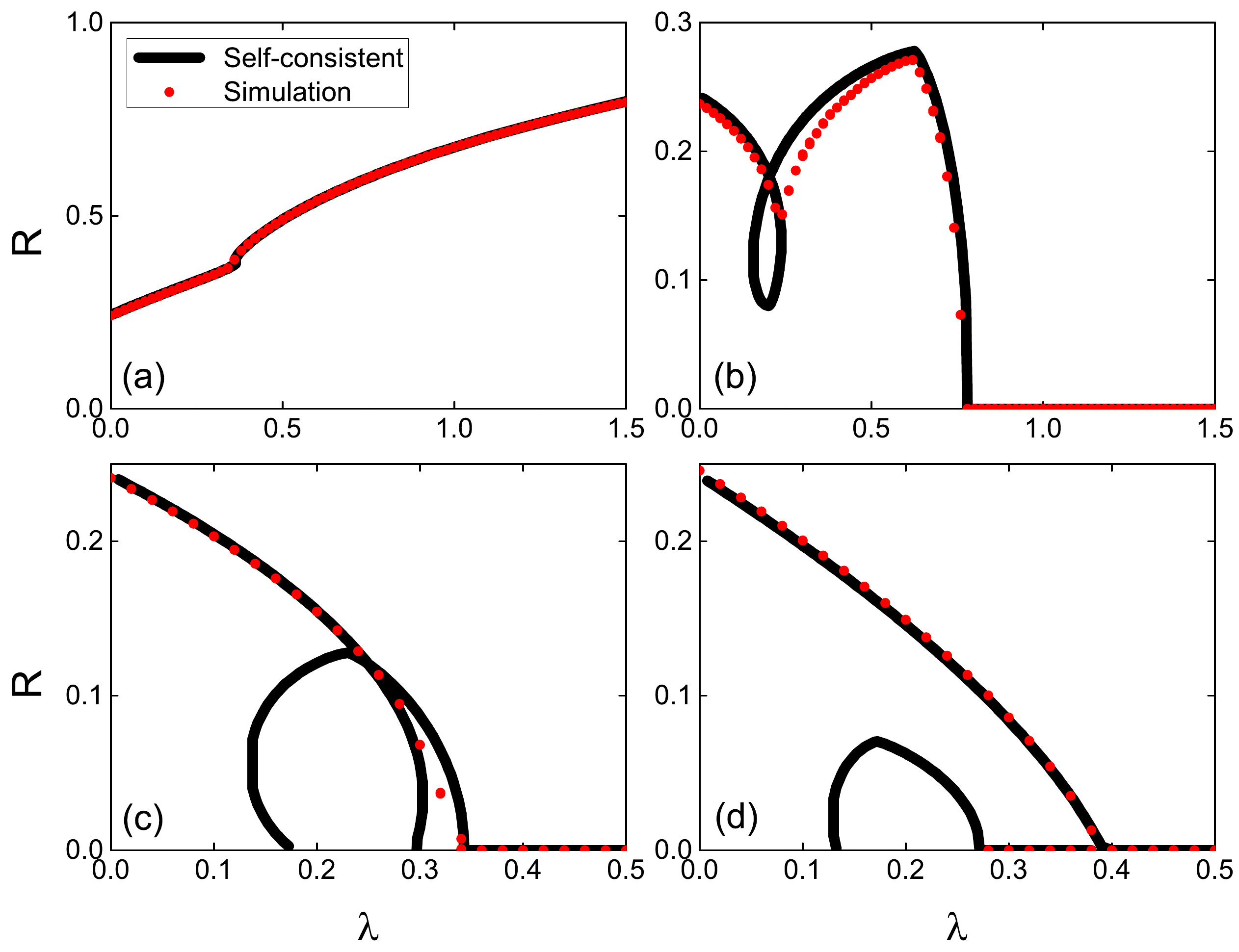}
\caption{\textbf{Direct numerical simulations for attractive second coupling.}
The global order parameter of traveling wave states as functions of $\lambda$. The parameters are $\Theta_2=0$, $\Delta=0.5$, and the values of $\Theta_1$ are (a): $\Theta_1=0.2\pi$; (b): $\Theta_1=0.6\pi$;
(c): $\Theta_1=0.75\pi$; (d): $\Theta_1=0.85\pi$. Black lines represent the self-consistent solution and red circles represent the direct simulation results. }
\label{fig:R-lambda2}
\end{figure}

\begin{figure}[!ht]
	\includegraphics[width=0.8\columnwidth]{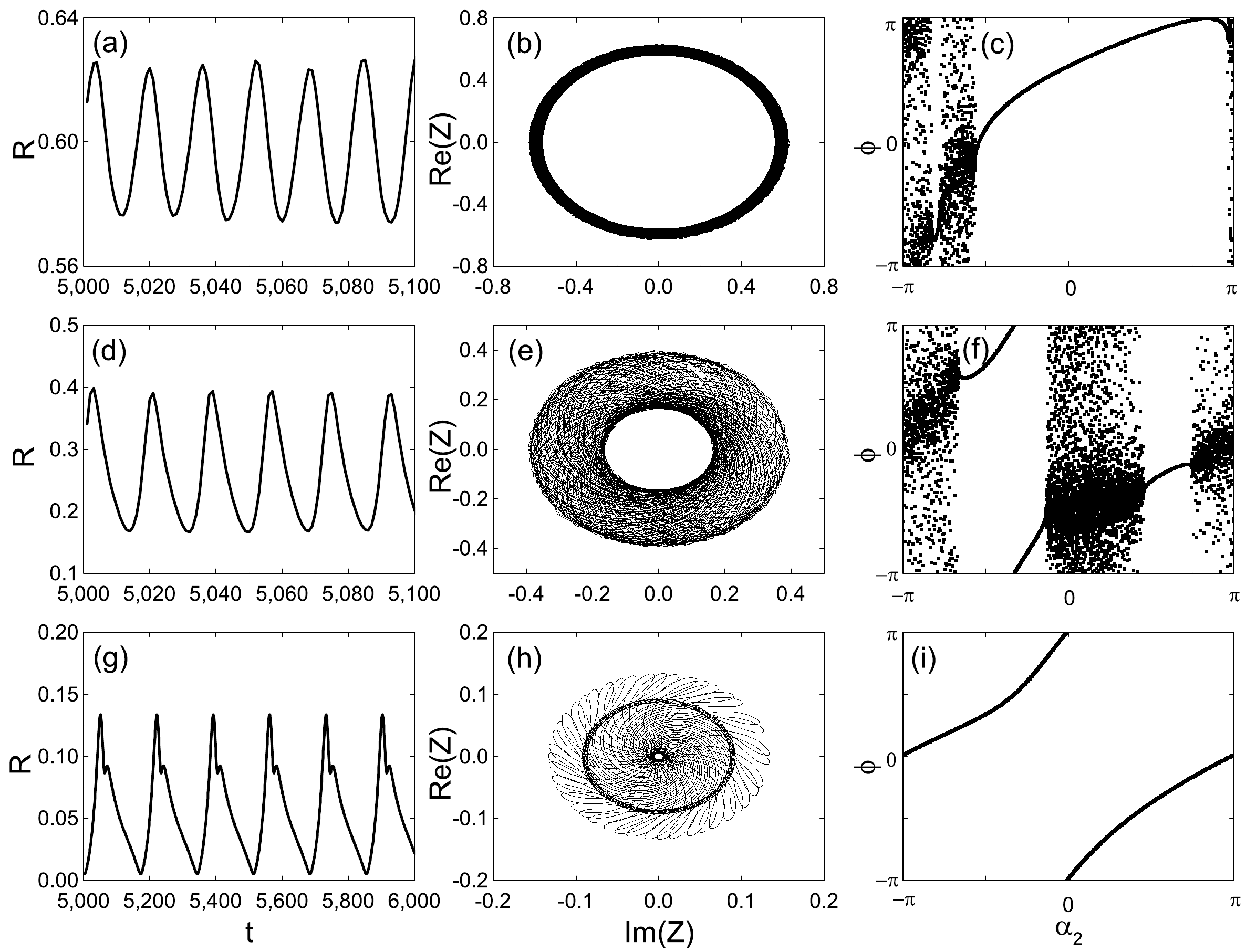} \caption{
		\textbf{Modulated synchronous states in direct numerical simulations.}
		Simulations for the parameters where no stable traveling waves exist,
		with $\Theta_2=0$ and $\Delta=0.5$. Panels (a), (d) and (g) show the time dependence
		of the global order parameter $R$,  after a long transient is discarded. Panels (b), (e) and (h) show the
		real and the imaginary part of the complex global order parameter $Z$. Panels (c), (f) and (i) show the snapshot of the phases of oscillators. The parameters are: in (a), (b) and (c), $\Theta_1=0.2\pi$, $\lambda=0.7$; in (d), (e) and (f),
		$\Theta_1=0.6\pi$, $\lambda=0.5$; in (g), (h) and (i), $\Theta_1=0.75\pi$, $\lambda=0.315$. }
	\label{fig:snapshot}
\end{figure}

Fig. \ref{fig:R-lambda2} shows the simulation results for traveling wave states depicted in Fig. \ref{fig:scsol3} above. The parameters are $\Theta_2=0$, $\Delta=0.5$ and $\Theta_1=$ (a): $0.2\pi$; (b): $0.6\pi$; (c): $0.75\pi$; (d): $0.85\pi$. The simulation results shows the normal, the anomalous,  and the
hybrid branches in Fig. \ref{fig:scsol3}, and also confirm our stability analysis. In the cases where no stable traveling waves exist, time-averaged values of the order parameter are shown. To reveal the modulated solutions, we show in Fig.~\ref{fig:snapshot}  the details of the system behavior for cases presented
in Fig.~\ref{fig:per} .  Panels (a), (d) and (g) show periodic in time evolution of the global order parameter $R$.
Panels (b), (e) and (h) demonstrate quasiperiodic rotations of the complex order parameter, confirming that the observed regimes are
modulated traveling waves. In panel (c), (f) and (i), one can see that the oscillators are partial locked. In fact, in panel (i) one cannot recognize the unlocked
pattern, because at the moment of time when the snapshot  has been performed, the local order parameter is everywhere close to one. Only during small time epochs over the period the unlocked region is visible (see discussion of ``blinking unlocking'' above.)

Next, we study systems with a small number oscillators. Here we focus mainly on situations of multistability, to check whether all the states
stable in the thermodynamic limit can be also observed for small populations.
Fig. 3 in the Supplementary Information shows the solutions for $N=100$ and the same parameters as
in Figs.~\ref{fig:scsol1},\ref{fig:R-lambda1} above
(parameters are $\Theta_1=0$, $\Delta=0.5$, and $\Theta_1=$ (a): $-0.4\pi$; (b): $-0.6\pi$; (c): $-0.9\pi$).
For each $\lambda$, we plot the final order parameters generated from 1500 random initial conditions.
Comparing with the results for a large ensemble $N=10000$  (Fig. \ref{fig:R-lambda1}), one can see that for the
cases of  $\Theta_2=-0.4\pi$ and $\Theta_2=-0.6\pi$, a small system shows the same behavior as
a large one. However, for the case of $\Theta_2=-0.9\pi$, we miss the anomalous solution in the small
system, only the normal branch is observed.
We conclude that the basin of this anomalous solution is rather small, so possibly finite-size fluctuations in a small system lead to a transition to the normal branch.

Fig. 4 in the Supplementary Information shows the probability of a finite-time
system to end up in the incoherent state, calculated for  50 realizations with
randomly chosen initial conditions.
Parameters are $\Theta_1=0$, $\Delta=0.5$ and $\lambda=0.2$ (see the green curve in Fig.~\ref{fig:scsol} for the
solutions in the thermodynamic limit).
The system sizes are (a): $N=10000$; (b): $N=5000$, (c): $N=1000$,
and (d): $N=500$; the  red lines show the linear stability boundaries according to Eq. \eqref{eq:stbo}. One can see that with the increasing of
system size, the probability of getting incoherent state are closer to 1 and the stability boundaries are also closer to the theoretical results.
For system size ensembles, the probability of getting incoherent state becomes lower.
This can be explained by the fluctuations of the order parameter, which increase with the decreasing system size, so the system
can jump out of the basin of the incoherent state due to the fluctuations. The nontrivial traveling wave state is relatively stable for these parameters and survives finite-size fluctuations.

\section*{Discussion}
In this paper, we studied the dynamics of a population of oscillators driven by two mean fields via
two different channels.  We have focused on the simplest setup, where all the oscillators have identical
frequencies, and the actions of the two mean fields differ by the coupling strengths and by the
distributions of coupling phase shifts.
The most close situation, previously discussed in the literature, is that of the standard globally coupled ensemble of oscillators (e.g., in the Kuramoto-Sakaguchi formulation) with an additional external feedback~\cite{Rosenblum-Pikovsky-04b,Rosenblum-Pikovsky-04c,Popovych-Hauptmann-Tass-05,Tukhlina-Rosenblum-Pikovsky-Kurths-07,Popovych-Tass-10}.  In these studies, however, it has been assumed that both channels
act with constant, non-distributed  coupling phases, what allowed for an essential reduction to a single-channel coupling.
Another situation close to the considered in this paper, is that of star-coupled oscillators
(i.e. all oscillators are coupled to one central, which then
represents one channel of global coupling)  with an additional
global mean field~\cite{Vlasov-Pikovsky-Macau-15}.

We have presented a general theory of traveling wave solutions in the thermodynamic limit, including
study of the stability in the Ott-Antonsen approximation. Mostly nontrivial states are those where
the two mean fields act oppositely, i.e. one tends to synchronize the oscillators and another desynchronizes them.
Here we have found different types of solutions: states where all oscillators are locked, states where
all oscillators are unlocked, and combined states where a part of oscillators are locked by driving fields, and another part
is unlocked (we stress that these states cannot be called chimeras, as here there is no symmetry in the population, because
the fields act on different oscillators with different phase shifts).

Stability of the found states has been studied using the Ott-Antonsen approach. The problem can be reduced to a
linear integral equation, which after discretization can be formulated as a matrix eigenvalue problem.
We have demonstrated existence of the essential continuous spectrum, two parts of which lie on the imaginary axis
(for unlocked oscillators) and on the real axis in the stable domain (for locked oscillators). Stability is determined by the discrete spectrum which in most cases can be found rather reliably.

By comparing amplitude behavior and stability of the found traveling wave solutions with the stability properties
of the incoherent state, we identified normal and anomalous branches of nontrivial solutions.
Normal branches bifurcate from the incoherent state when the latter becomes unstable, typically in
a subcritical way. The corresponding transitions are hysteretic and discontinuous, which is similar to
recent findings of the discontinuous transition in other synchronization
setups~\cite{Gardenes-Gomes-Arenas-Moreno-11,Leyva-etal-12,Zhang-Hu-Kurths-Liu-2013,Zhang-Boccaletti-Guan-Liu-2015,Zhang-Hu-Kurths-Liu-2014}.
Additionally, we have found that some branches of solutions appear with vanishing order parameter at points
where no change of stability of the incoherent state occurs. These anomalous branches thus do not bifurcate
from the incoherent state. Some parts of the
anomalous branches may be stable, and they are observed in the simulations
of the Ott-Antonsen equations describing the thermodynamic limit. However, in
simulations of small finite ensembles the
population ``prefers'' more robust normal stable branches. Furthermore, we have found
hybrid branches, which combine
properties of the normal and anomalous ones: such a branch starts from a vanishing order
parameter at a point
where the incoherent state becomes unstable, but ends at a point where stability of the
incoherent state does not change.

Additionally, we have found situations where no stable traveling waves exist. Here numerical
simulations, both in the
thermodynamic limit and in finite ensembles, demonstrated modulated traveling waves with periodically
in time pulsating order parameter. Remarkably, in some cases such a wave looks like blinking
unlocking: there are epochs when all
oscillators look like being locked, while in other epochs one can clearly see a wide distribution of the phases.

 Actually, in the situation which the oscillators are coupled by three or more mean fields, and all the channels act with constant. The system can be reduce to our framework by summing the attractive/repulsive couplings into resultant attractive/repulsive mean field, and then it can be studied base on our results. But if the coupling channels have more complicated forms, and can not be summed into resultant mean fields, then one needs to develop a specific approach for the situation. 

In conclusion, the competition between two coupling mean fields can make the system behavior much more
complicated than
the case with only one channel of global coupling. Further
study in this direction would help to understand nontrivial synchronization regimes in complex systems in
nature, such as coupled cellular systems and neuronal systems.

\section*{Methods}

\subsection*{Linear stability of the incoherent state}
In this section we perform a linear stability analysis of the fully incoherent
(i.e. with a uniform distribution of phases) state of system~\eqref{eq:bm4} in the thermodynamic limit
of infinite number of oscillators. In this limit one starts with a continuity
equation for the density function $\rho (\theta, t, \alpha)$
\begin{equation}
\label{density}
\frac{\partial \rho}{\partial t}+\frac{\partial (\rho \upsilon)}{\partial \phi}=0\;,
\end{equation}
with the velocity $
\upsilon =R\lambda\sin(\Psi-\phi+\Theta_1)+R\sin(\Psi-\phi+\Theta_2+\alpha)$ and the mean field
\[
Re^{i\Psi}=\iint_0^{2\pi}d\phi\,d\alpha\, e^{i\phi}\rho(\phi,t,\alpha)g(\alpha)\;.
\]
Suppose there is a small perturbation from the incoherent state $\rho=(2\pi)^{-1}$, i.e.,
 \begin{equation}
\label{perturbation}
\rho (\phi , t, \alpha)=\frac{1}{2\pi}+\epsilon \eta (\phi , t, \alpha)
\end{equation}
with $\epsilon\ll1$.  Expanding the perturbation $\eta$ in the Fourier series
 \begin{equation}
\label{Fourier}
\eta (\phi , t , \alpha)=c(\alpha, t)e^{i\phi}+c^{*}(\alpha, t)e^{-i\phi}+c^{\bot}(\alpha, t)\;,
\end{equation}
where $c^{\bot}(\alpha, t)$ is the sum of higher harmonics, we can represent the mean field
as
\[
Re^{i\Psi}=\epsilon 2\pi \int_{0}^{2\pi}c^{*}(\alpha, t)g(\alpha)d\alpha
\]
Substituting these relations into~\eqref{density} and separating the Fourier modes, we get, to the first order in $\epsilon$,
the following equation
for the evolution of $c$:
\begin{equation}
\label{eq1}
\frac{\partial c(\alpha, t)}{\partial t}=\frac{1}{2}[\lambda e^{-i\Theta_1}\int_{0}^{2\pi}c(x,t)g(x)dx+e^{-i(\Theta_2+\alpha)}\int_{0}^{2\pi}c(x,t)g(x)dx]\;.
\end{equation}
With the exponential in time ansatz $c(\alpha,t)=e^{\gamma t} b(\alpha)$ Eq.~\eqref{eq1} takes the form
 \begin{equation}
\label{eq2}
\gamma b(\alpha)=\frac{1}{2}[\lambda e^{-i\Theta_1}\int_{0}^{2\pi}b(x)g(x)dx+e^{-i(\Theta_2+\alpha)}\int_{0}^{2\pi}b(x)g(x)dx]
\end{equation}
Equation \eqref{eq2} can be solved in a self-consistent way.  We define
$A=\frac{1}{2}\int_{0}^{2\pi}b(x)g(x)dx$, so that
$b(\alpha)$ can be expressed as $b(\alpha)=\frac{1}{2\gamma}[\lambda e^{-i\Theta_1}
A+e^{-i(\Theta_2+\alpha)}A]$. Thus we obtain from Eq.~\eqref{eq2}
 \begin{equation}
\label{eq3}
1=\frac{\lambda e^{-i\Theta_1}}{4\gamma}+\frac{e^{-i\Theta_2}}{4\gamma}\int_{0}{2\pi}e^{-i\alpha}g(\alpha)d\alpha
\end{equation}
For a particular case of the von Mises distribution~\eqref{eq:vmd}
the integral can be explicitly calculated and we obtain
\[
\gamma=\frac{\lambda e^{-i\Theta_1}}{4}+\frac{e^{-i\Theta_2}I_1(\Delta)}{4 I_0(\Delta)}
\]
Stability is determined by the real part of $\gamma$, so the critical values of the
parameters, separating stable and unstable incoherent state,
can be obtained from the equation
\begin{equation}
\text{Re}(\gamma)=\frac{\lambda \cos \Theta_1}{4}+\frac{I_1(\Delta)\cos \Theta_2}{4I_0(\Delta)}=0
\label{eq:stbo}
\end{equation}

\subsection*{Self-consistent approach}
\label{sec:urs:sca}
We seek for partially synchronous solutions in the model described by Eqs.~\eqref{eq:bm3} in the form of uniformly
rotating states, with some
frequency $\Omega$,  to be defined in the procedure.  In terms of the distribution density $\rho(\phi,\alpha,t)$ these solutions are traveling waves. It is convenient to transform all the variables to the
rotating frame, where we then will look for stationary solutions of the distribution of the phases.
Furthermore, because we treat the problem
in the thermodynamic limit, it is suitable to parametrize the phases by the value of parameter $\alpha$ (we will write it as a lower index for the phase variables).
We thus introduce new phases
\[
\psi=\theta-\Psi-\Theta_1
\]
and assume that $\dot\Psi=\Omega$. In these variables the ensemble driven by two mean fields
\eqref{eq:bm4} can be written as
\be
\dot\psi_\alpha=-\Omega-R\lambda\sin\psi_\alpha-R\sin(\psi_\alpha+\beta -\alpha),\qquad
Re^{-i\Theta_1}=\langle e^{i\psi}\rangle=\int_0^{2\pi} d\alpha g(\alpha)\exp[i\psi_\alpha]
\label{eq:2}
\ee
where $\beta=\Theta_1-\Theta_2$.
It is convenient to introduce auxiliary parameters $a,r,s$ according to
\be
a=\frac{R}{\Omega},\quad r^2=a^2\lambda^2+2a^2\lambda\cos(- \beta+\alpha)+a^2,\qquad
s=\text{atan2}({ a}\sin(-\beta+\alpha),a\lambda+a\cos(-\beta+\alpha))\;.
\label{eq:parms}
\ee
Then
\be
\dot\psi_\alpha=
\Omega(-1-r\sin(\psi_\alpha-s))
\label{eq:psieq}
\ee

One can see that the essential parameters for the dynamics of oscillators in \eqref{eq:psieq}
are $r$ and $s$, which depend
explicitly on $a,\lambda,\beta,\alpha$. This suggests the following strategy to find the solutions of the model:
\begin{enumerate}
\item For fixed parameters $a,\lambda,\beta,\alpha$ one finds  a stationary density of phases $w(\psi_\alpha|a,\lambda,\beta,\alpha)$
and the corresponding average
\begin{equation}
z(a,\lambda,\beta,\alpha)=\langle e^{i\psi_\alpha}\rangle=\int_0^{2\pi}d\psi_\alpha \exp[i\psi_\alpha] w(\psi_\alpha|a,\lambda,\beta,\alpha)
\end{equation}
\item After that one uses the value of $z$ to find the average over the distribution of $\alpha$
\[\langle e^{i\psi}\rangle=
\int_0^{2\pi}d\alpha g(\alpha)z(a,\lambda,\beta,\alpha) =F(a,\lambda,\beta)\exp[iQ(a,\lambda,\beta)]\;.
\]
\item  Applying \eqref{eq:2} and \eqref{eq:parms} one then obtains the parameters $\Theta_{1,2}$ as well as the main
characterizations of the dynamics, the order parameter $R$ and the frequency $\Omega$, in a parametric form
as functions of the introduced auxiliary parameters $a,\beta$:
\be
\begin{gathered}
R=F(a,\lambda,\beta),\qquad \Theta_1=-Q(a,\lambda,\beta),\\
\Theta_2=-Q(a,\lambda,\beta)-\beta,\qquad \Omega=\frac{F(a,\lambda,\beta)}{a}\;.
\end{gathered}
\label{eq:rf}
\ee
\end{enumerate}

We now present the steps in this procedure in details. The first goal is to find a distribution
of the phases governed by
Eq.~\eqref{eq:psieq}. There are two types of possible regimes for the phase: the locked state for $r>1$ and the rotating state
for $r<1$. The locked state has a definite value of the phase $\psi=\psi_0$, thus the distribution
is the delta-function and $z_l=\exp[i\psi_0]$ (here index $l$ denotes locked states). We have to choose the stable locked state, therefore the value of $\psi_0$
depends also on the sign of $\Omega$, i.e. on the
sign of parameter $a$:
\be
z_l=\begin{cases}e^{is}(\sqrt{1-(1/r)^2}-i/r)& \text{ for }\Omega>0\\
e^{is}(-\sqrt{1-(1/r)^2}-i/r)& \text{ for }\Omega<0
\end{cases}
\label{eq:zl}
\ee
For $r<1$ there are no locked states and the phases rotate.
 Here the stationary distribution is just inverse proportional to the velocity
\[
w(\psi)=\frac{C}{|\dot\psi|}=\frac{C}{1+r\sin(\psi-s)}
\]
where $C$ is the normalization constant. Using standard integrals we get
\be
w(\psi)=\frac{\sqrt{1-r^2}}{2\pi(1+r\sin(\psi-s))},\qquad z_r=ie^{is}\frac{-1+\sqrt{1-r^2}}{r}
\label{eq:zr}
\ee
(Here index $r$ denotes rotating states). Combining Eqs.~\eqref{eq:zl},\eqref{eq:zr}
we get the final expression for the parameters $F,Q$:
\be
\langle e^{i\psi}\rangle=Fe^{iQ}=\int_0^{2\pi}d\alpha z_{l} g(\alpha) +\int_0^{2\pi}d\alpha z_{r} g(\alpha),\qquad
g(\alpha)=\frac{e^{\Delta\cos\alpha}}{2\pi I_0(\Delta)}
\label{eq:fq}
\ee
Additionally, we can calculate the portion of the locked oscillators $P_l$:
\be
P_l(a,\lambda,\beta)=\int_0^{2\pi}d\alpha H(\alpha) g(\alpha)
\label{eq:plock}
\ee
where
\[
H(\alpha)=\begin{cases} 1&\text{ if }r>1,\\
0&\text{ if }r<1,
\end{cases}
\]
is the indicator function for the locked states. We call $P_l$ the locking parameter.
In the standard Kuramoto model, $P_l$ is proportional to the mean field
amplitude, but in our case we will see a nontrivial behavior of $P_l$: in some cases
all the oscillators are locked, i.e. $P_l=1$, while
the order parameter $R$ is rather small.

\subsection*{Reformulation in terms of Ott-Antonsen equations}
Generally, to study stability of the stationary rotating solutions described above, we need to analyze generic perturbations
of the equation for the density \eqref{density}. We however restrict ourselves to a class of perturbations lying on the
so-called Ott-Antonsen manifold~\cite{Ott-Antonsen-08}.
The coupling in our model has a pure sin form, therefore the Ott-Antonsen (OA) ansatz leading to an integral equation
for the local order parameter $z(\alpha,t)=\langle e^{i\psi_\alpha}\rangle$ is possible.
In this Ansatz one represents the density as $\rho(\psi,\alpha,t)=(2\pi)^{-1}[1+\sum_m (z^m e^{-im\psi}+(z^*)^{m}e^{im\psi})]$.
Then one applies this to Eq.~\eqref{eq:2} and obtains (see~\cite{Ott-Antonsen-08,Pikovsky-Rosenblum-15} for details)
\be
\begin{gathered}
\dot z(\alpha,t)=-i\Omega z+\frac{1}{2}\left(\left(Z \lambda e^{i\Theta_1}+Ze^{i\Theta_2}e^{i\alpha}\right)-
\left(Z^* \lambda e^{-i\Theta_1}+Z^*e^{-i\Theta_2}e^{-i\alpha}\right)z^2(\alpha)\right),\\
Z=\int_0^{2\pi} g(\alpha) z(\alpha)\;.
\label{eq:oac}
\end{gathered}
\ee

Because the found solutions are steady states of system~\eqref{eq:oac}, the linearization is straightforward. Furthermore,
we discretize the resulting linear integral equations using a discrete (size $L$) representation
of the parameter $\alpha$ on the circle:
$\alpha_k=\alpha_0+2\pi k/L$, $k=0,\ldots,L-1$. Here the offset $0\leq \alpha_0<2\pi/L $ is a
free parameter. It can be used to distinguish the
continuous and the discrete parts of the spectrum (cf. Ref.~\cite{Smirnov_et-al-16}).

\subsection*{Solutions with a vanishing order parameter}
\label{app:vop}
Here we analyze the limit $R\to 0$ of the constructed above traveling wave solutions.
Because, according to~\eqref{eq:rf},  $R=\Omega a$, there are two possibilities for order parameter
$R$ to vanish: $a=0$ and $\Omega=0$.

\subsubsection{Case $a\to 0$}

For $a\to 0$ we have also $r\to 0$.
Equation \eqref{eq:zr} in the limit $r\to 0$ reduces to
\[
\langle e^{i\psi}\rangle_r=ie^{is}\frac{-1+\sqrt{1-r^2}}{r}\approx -\frac{r}{2}ie^{is}
\]
We use
\[
re^{is}=a\lambda+a\cos(\alpha-\beta)+i a\sin(\alpha-\beta)
\]
and get from \eqref{eq:fq}
\be
\begin{gathered}
Fe^{iQ}=-\frac{i}{2} \int_0^{2\pi} d\alpha g(\alpha) [a\lambda+a\cos(\alpha-\beta)+i a\sin(\alpha-\beta)]=\\
=-\frac{i}{2}[\lambda a+ a e^{-i\beta}\int_0^{2\pi} d\alpha g(\alpha)e^{i\alpha}]=
-\frac{i}{2}[\lambda a+ a e^{-i\beta}\frac{I_1(\Delta)}{I_0(\Delta)}]
\end{gathered}
\ee
We substitute here $F=R=\Omega a$, $Q=-\Theta_1$ and $\beta=\Theta_1-\Theta_2$ and obtain in
 the  limit $a\to 0$
\[
2 \Omega =-[\lambda (i\cos \Theta_1-\sin \Theta_2)+(i\cos \Theta_2-\sin \Theta_2) \frac{I_1(\Delta)}{I_0(\Delta)}]
\]
Because $\Omega$ is real, the condition
\[
\lambda \cos \Theta_1+\cos \Theta_2\frac{I_1(\Delta)}{I_0(\Delta)}=0
\]
should be fulfilled. This is exactly the condition for the  linear stability border~\eqref{eq:stbo}.

\subsubsection{Case $\Omega= 0$}
In this case $F=0$ and according to \eqref{eq:fq}
\be
0=\int_0^{2\pi}d\alpha\langle e^{i\psi}\rangle_{r,l} g(\alpha)
\label{eq:aav1}
\ee
This relation defines two real equations, i.e. two constrains on parameters $a,\beta,\lambda$. This means that  there
may be a curve (or a set of curves) $\beta(\lambda)$ for which condition \eqref{eq:aav1} is fulfilled, so that solutions
with vanishing order parameter $R=0$ exist on this set. This is exactly the
curve of anomalous solutions with vanishing $R$ presented in Fig~\ref{fig:linst}.

\newpage
{\bf Acknowledgments}
The work was supported by ITN COSMOS (funded by
the European Union's Horizon 2020 research and innovation
programme under the Marie Sklodowska-Curie grant agreement No 642563), and by the NNSF of China
under Grant No. 11375066.
The analysis of the OA system was
supported by the Russian Science Foundation
(Project No. 14-12-00811).

{\bf Author Contributions}
X.Z. and A.P. conceived the research project.
X.Z. and A.P. performed research. X.Z., A.P and Z.L. analyzed
the results. X.Z. and A.P. wrote the paper. All Authors reviewed the Manuscript.

{\bf Competing Interests}
The authors declare no competing financial interests.

{\bf Correspondence} Correspondence and requests for materials
should be addressed to X.Z. ($zxy\_822@126.com$).

\newpage
\section*{Supplementary Information}

\begin{figure}[!hbt]
\includegraphics[width=0.6\columnwidth]{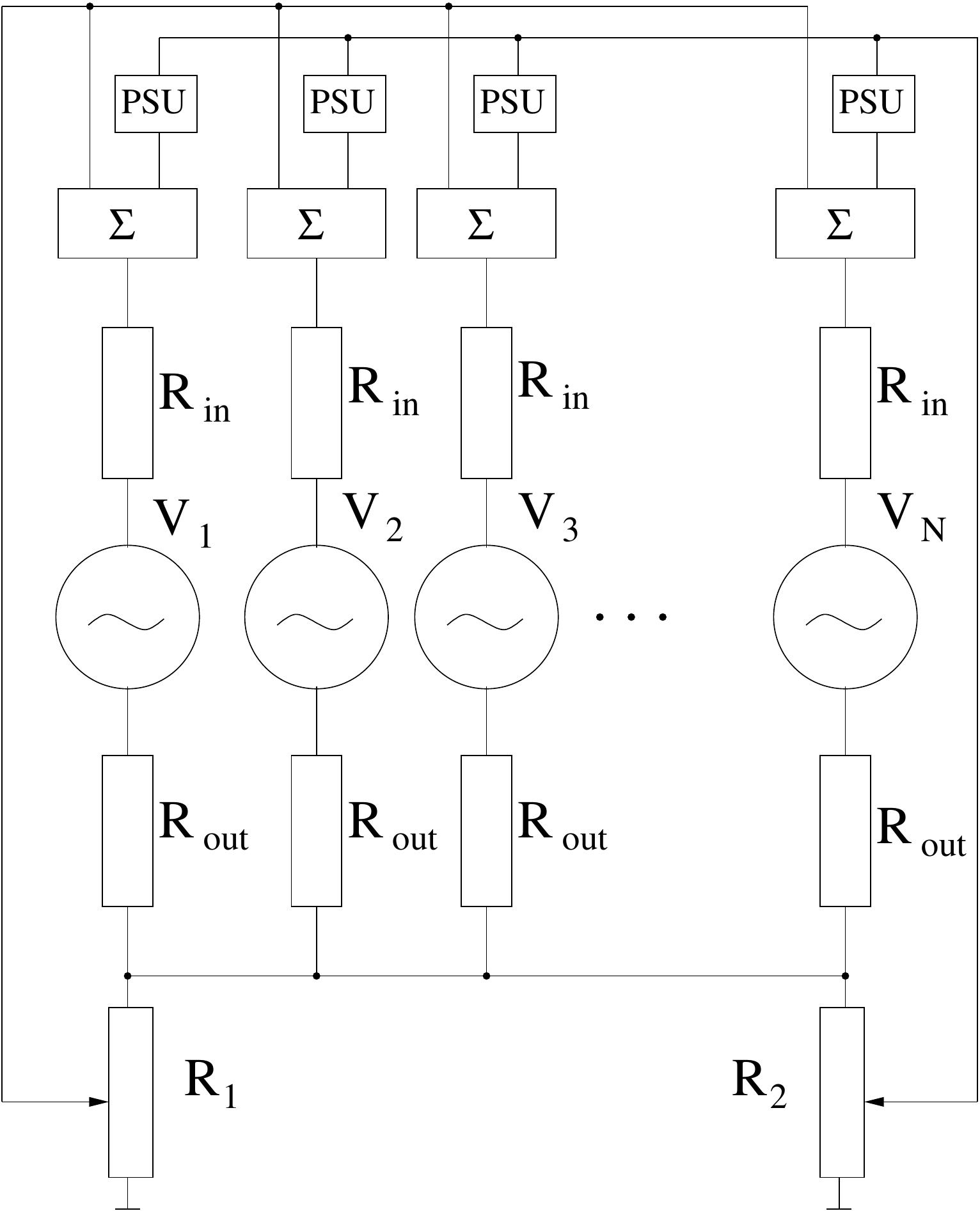} \caption{
\textbf{A possible electronic circuit  implementing two mean field coupling.}  It consists of $N$ possibly identical
(within possible accuracy of realization) Wien-bridge selv-sustained oscillators. Their equations are
equivalent to the van der Pol equation. Two global couplings are organized via the common resistive loads $R_1$ and $R_2$.
The voltages from these loads are fed back to all oscillators. One voltage is fed back directly, while the other voltage is fed back via 
phase shifts units. Thus, this scheme implements the basic model studied in this paper.}
\label{Fig:scheme}
\end{figure}

\begin{figure}[!hbt]
\includegraphics[width=0.6\columnwidth]{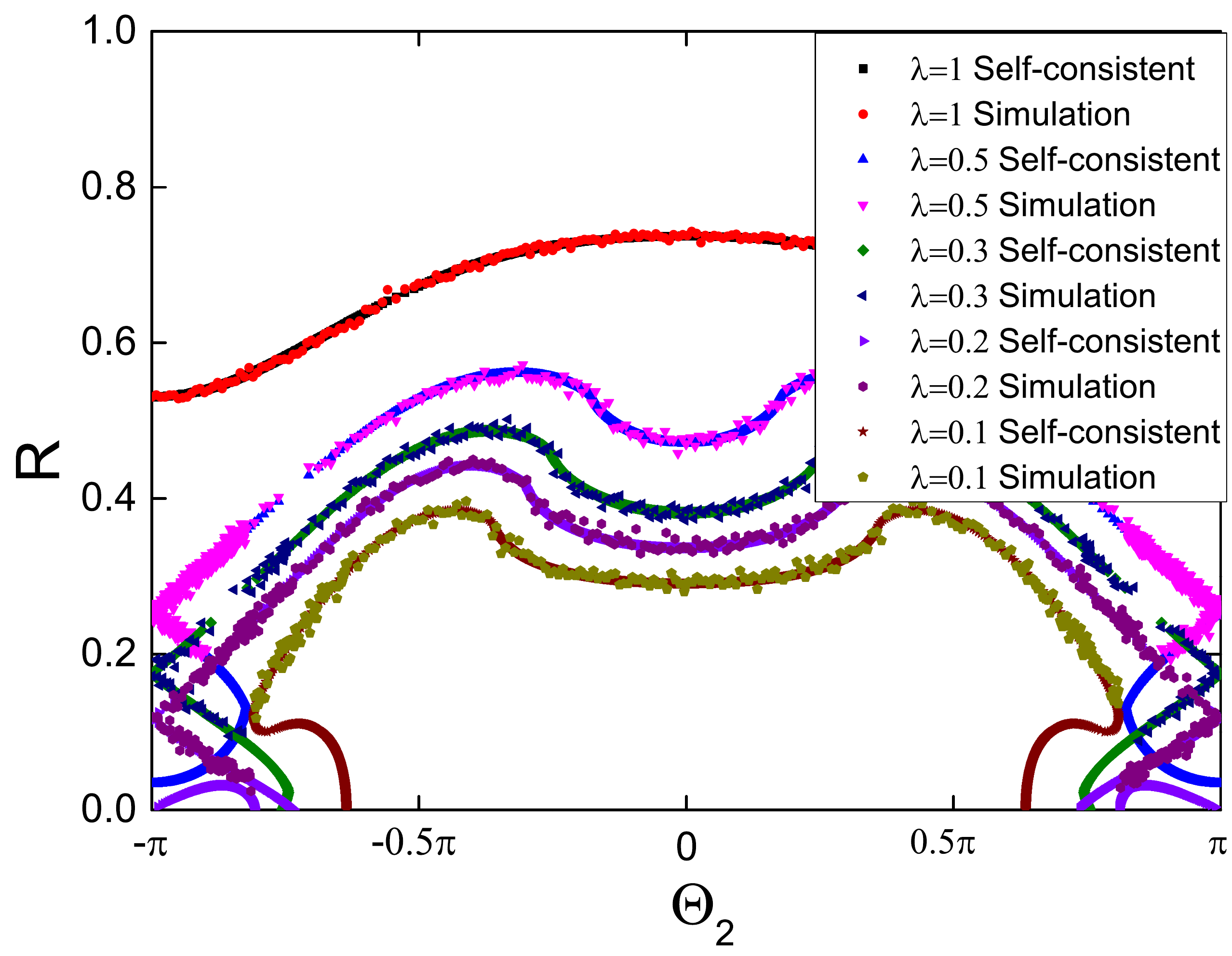} \caption{
\textbf{Compare the self-consistent solution and direct simulation results}, the parameters are the same with Fig. 3 in the main body. }
\label{Fig:SC-SI-Q1}
\end{figure}

\begin{figure}[!h]
\includegraphics[width=0.6\columnwidth]{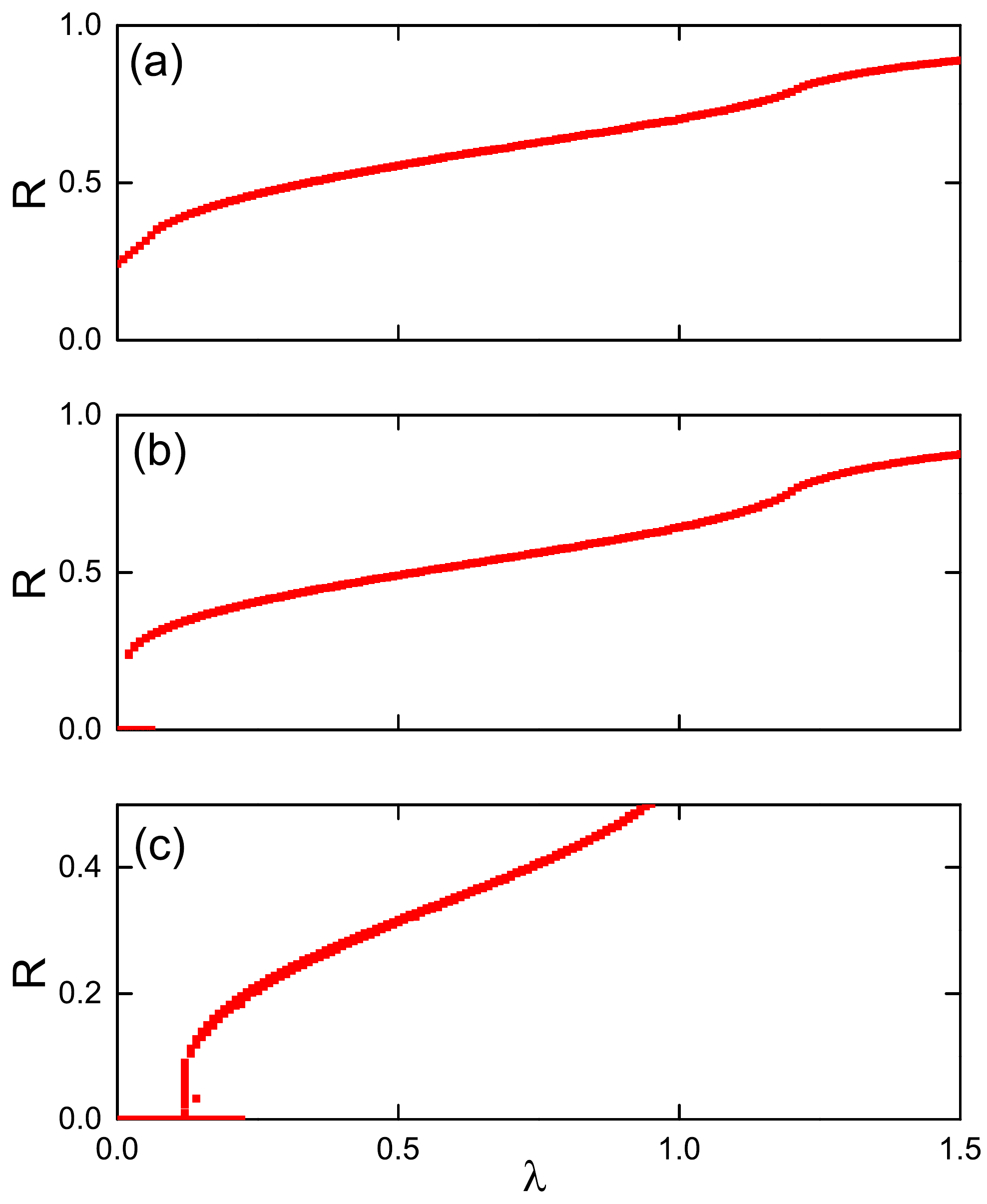} \caption{
\textbf{The global order parameter of traveling wave states for $\Theta_1=0$ and $\Delta=0.5$, as functions of $\lambda$.} The values
of $\Theta_2$ are (a): $\Theta_2=-0.4\pi$; (b): $\Theta_2=-0.6\pi$; and (c): $\Theta_2=-0.9\pi$. Results are from simulations
with a N=100 system and for each $\lambda$, we plot the final order parameter generated from 1500 random initial conditions.}
\label{fig:R-lambda1-100}
\end{figure}

\begin{figure}[!h]
\includegraphics[width=0.6\columnwidth]{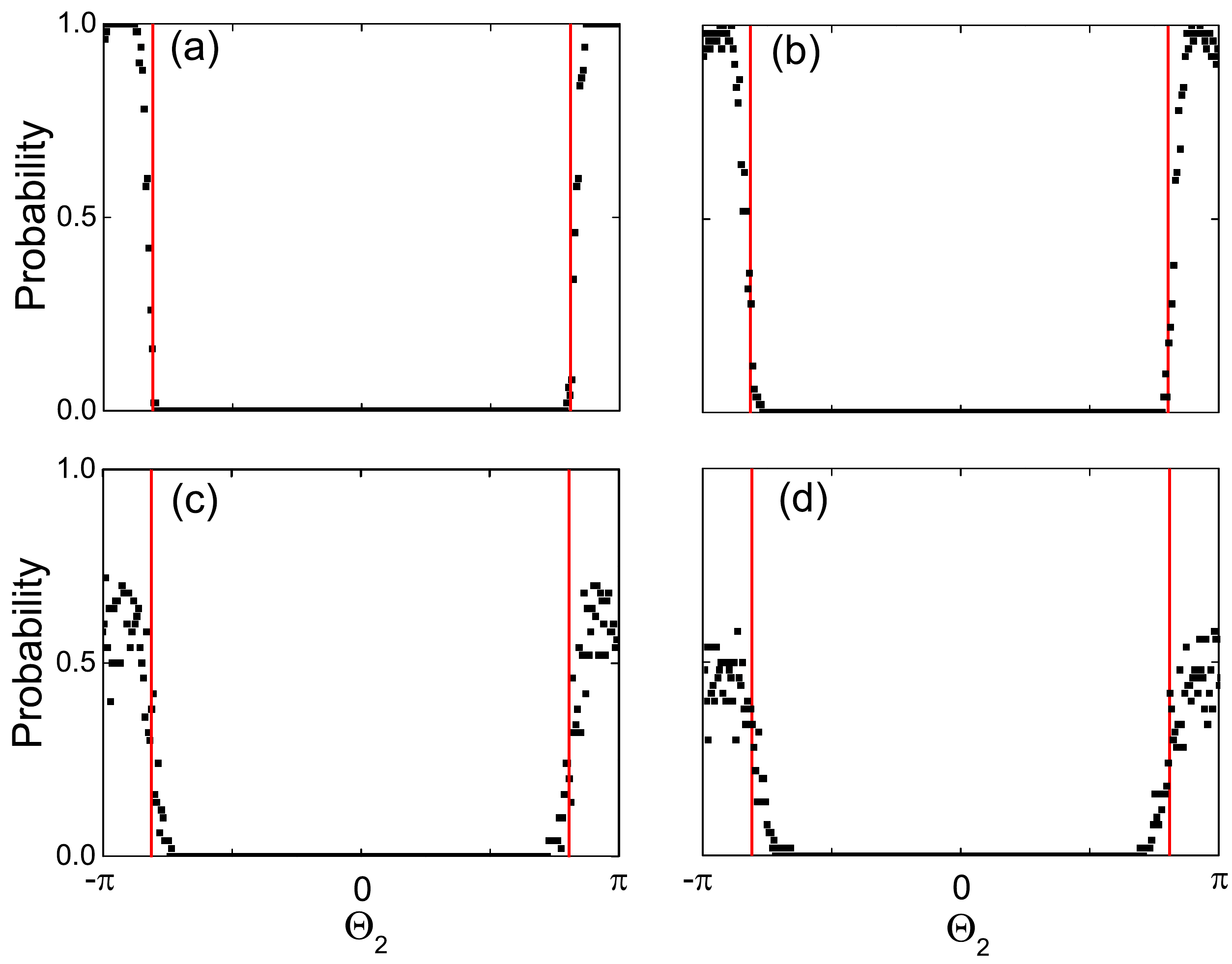} \caption{
\textbf{Probability of the system ending up with incoherent state in 50 realizations with random initial conditions.} Parameters are $\Theta_1=0$, $\Delta=0.5$ and $\lambda=0.2$. The system sizes are (a): N=10000; (b): N=5000, (c): N=1000 and (d): N=500. Red lines show the linear stability boundaries according to Eq. (15) }
\label{fig:stability-size}
\end{figure}

\end{document}